\documentclass[10pt]{iopart}

\usepackage{harvard}
\usepackage{graphicx}
\usepackage{undertilde}
\usepackage{xcolor}

\usepackage{iopams}
\citationmode{abbr}
  
\begin{document}

\title[Validation of a model for an ionic electro-active polymer in the static case]{Validation of a model for an ionic electro-active polymer in the static case}

\author{M. Tixier$^1$ \& J. Pouget$^2$}

\address{$^1$ Laboratoire de Math\'ematiques de Versailles (LMV), UMR 8100, Universit\'e de Versailles Saint Quentin, 45, avenue des Etats-Unis, F-78035 Versailles, France}
\address{$^2$ Sorbonne Universit\'e, CNRS, Institut Jean le Rond d'Alembert, UMR 7190, F-75005 Paris, France}
\ead{\mailto{mireille.tixier@uvsq.fr}, \mailto{pouget@lmm.jussieu.fr}}
\vspace{10pt}
\begin{indented}
\item[]April 2020
\end{indented}

\begin{abstract}
IPMCs consist of a Nafion$^{\mbox{\scriptsize{\textregistered}}}$ ionic polymer film coated on both sides with a thin layer of metallic electrodes. The polymer completely dissociates when it is saturated with water, releasing small cations while anions remain bound to the polymer chains. When this strip is subject to an orthogonal electric field, the cations migrate towards the negative electrode, carrying water away by osmosis. This leads to the bending of the strip.
We have previously published a modelling of this system based on the thermodynamics of irreversible processes. In this paper, we use this model to simulate numerically the bending of a strip. Since the amplitude of the deflection is large, we use a beam model in large displacements. In addition, the material permittivity may increase with ion concentration. We therefore test three permittivity models. We plot the profiles of the cations concentration, pressure, electric potential and induction, and we study the influence of the strip geometry on the tip displacement and on the blocking force. The results we obtain are in good agreement with the experimental data published in the literature. The variation of these quantities with the imposed electric potential allow us to discriminate between the three models.
\end{abstract}

\vspace{2pc}

\newline
\newline
published in \textit{Smart Materials and Structures} \textbf{29}(085019), 2020

https://doi.org/10.1088/1361-665X/ab8fca

\ioptwocol

\section{Introduction}
The present work proposes a study of a thin strip of electro-active polymer (EAP). More precisely, the study focuses on  Ionic Polymer-Metal Composite (IPMC) belonging to the ionic class. We investigate responses of the EAP subject to an applied difference of electric potential on the metallic electrodes and an applied punctual  force at the tip of the strip. In  previous works we have constructed step by step a micro-mechanical model to establish the conservation laws for electro-active polymers (Tixier \& Pouget 2014). Following this work, the constitutive equations were deduced from the hypothesis of local thermodynamics  equilibrium and the Gibbs relation using the thermodynamics of irreversible processes (Tixier \& Pouget 2016). The present study follows on from the spirit of the previous works and we want to characterize the behavior of a thin blade, especially the chemical, electrical and mechanical parameters under electro-mechanical loading.

\noindent
The behavior of electro-chemical-mechanical interactions of EAP is of great interest for research and engineering advanced technology. Simply, we can say that an EAP is a polymer exhibiting a mechanical response, such a stretching, contracting or bending for example, when subject to an electric field. Conversely the EAP can produce electric energy in response to a mechanical load (Shahinpoor et al. 1998, Shahinpoor \& Kim 2001, Bar-Cohen 2005, Pugal et al. 2010, Park et al. 2010). This particular property is highly attractive for applications. For instance, we can quote biomimetic devices (robotics, bio-inspired underwater robots such as fishes (Chen 2017), haptic actuators (artificial skin, tactile displays or artificial muscle (Deole et al. 2008, Matysek et al. 2009, Bar-Cohen 2005) and this material is an excellent candidate for energy harvesting (Farinholt et al. 2009, Tiwari et al. 2008, Aureli et al. 2010, Jean-Mistral et al. 2010, Cha et al. 2013). In addition promising applications to micro-mechanical systems (MEMS) at the sub-micron scale are also expected in medical engineering for accurate medical control or investigation, for instance Fang et al. (2007) and Chikhaoui et al. (2018). \\

\noindent
Electro-active polymers can be categorized into two main groups depending on their activation mechanisms. The first one is electronic electro-active polymers and dielectrics which are subject to Coulomb force ; their volume change is due to the application of an electric field. For instance dielectric elastomers belongs to this category as well as piezoelectric and electrostrictive polymers. The second group of EAP is the ionic electro-active polymers. The latter are driven by the displacement of ions inside the material. The polyelectrolyte gels, ionic polymer-metal composites, conductive polymers are such media. One of the advantages of the these polymers is that they can be activated by very low difference of electric potential of about $1 - 5$ Volts. However, they can only be operated within electrolyte medium. 

Most efforts have been devoted on electronic conducting polymers (E.C.P.) belonging to the ionic class in order to improve strain, output forces and response times. 
These polymers are used for multilayer polymer composite combined with a polymer which can be considered as ion reservoir to improve ion transfer. Trilayer actuators made of solid polymer electrolyte film sandwich between two electronic conducting polymers have been reported in Festin et al. (2014) and Nguyen et al. (2018). One of the advantage of these electro-active polymers is that they can operate in air and they are good candidates for biomimetic devices. Efforts have been conducted to increase the integrating biocompatible conducting polymers into continuum micro-robots. Developments of such biocompatible conducting polymers with the aim at designing accurate position control for the trajectory of the telescopic soft robots have been studied by Chikhaoui et al. (2018). \\

\noindent
Modeling EAP is not an easy task, especially the model must include electro-mechanical and chemical-electric couplings of ion transport, electric field action and elastic deformation. Different kinds of approach have been proposed in the literature according to the underlying physics of chemical activation (Shahinpoor \& Kim 2001, Brunetto et al. 2008, Deole et al. 2008, Bahramzadeh \& Shahinpoor 2014).
An instructive and comprehensive review paper devoted to IPMC has been proposed  by Jo et al. (2013). The authors present the chemico-physical 
mechanisms involved in IPMCs.  As a complement to the previous references, a set of interesting studies has been devoted to the modeling of EAP. Among them, Wallmersperger et al. (2009) 
develope a thermodynamically based mechanical model involving the chemico-electric transport phenomenon. Numerical simulations are proposed for a strip of EAP made of 
Nafion$^{\mbox{\scriptsize{\textregistered}}}$ $117~Li^{+}$. The authors deduce the profiles, within the strip thickness, of the electric charge density, electric potential, electric field and strain.  Nardinocchi et al. (2011) deduce a model based on the 3-D theory of linear elasticity. The thermodynamics allows them to introduce chemo-electro-mechanical coupling and they deduce the 
constitutive equations of the material, especially a Nernst-Planck like equation is deduced. Their study continues with numerical illustrations for a thin strip of EAP. 
Moreover, time evolution at low frequency is proposed as well. It is noteworthy that these papers are partly, in their spirit, rather close to the present 
model. Nevertheless, their results depend strongly on adjustable parameters to fit the experimental results. In their model, the Fourier law, Darcy law and the generalized rheological 
constitutive equation are not presented. Moreover, using their adjustable phenomenological parameters the authors deduce a dielectric permittivity greater than ours.

A continuum approach for multiphasic materials has been investigated by Bluhm et al. (2016). The authors 
write down the conservation laws and the entropy inequality based on the theory of porous media for an arbitrary number of individual constituents. In spite of the rather thorough  
investigation, the authors do not compare their results with those  proposed in the literature. However, their constitutive laws are related to the material constituents and not to the macroscopic medium, and their coefficient are not numerically evaluated. \\ 

The present model is mainly based on an averaging method of the microscopic description of phases in order to 
deduce the macroscopic behavior of the material. Thanks to this approach, the constitutive coefficients for the whole material are computed with the help of  the microscopic 
physico-chemical properties of the constituents (volume fractions, mass densities, chemical potentials, electric charges, etc.) and physical meanings are identified. \\

\noindent
An important polymer property that we would like to address is the dependency of the dielectric permittivity with the cation concentration in the polymer. As matter of fact, the dielectric constant of the polymer is not absolutely homogeneous within the strip thickness. Accordingly, it seems that the ionic transport when a difference of electric potential is applied to the polymer strongly modifies the dielectric constant along the thickness direction of the blade. This question will be examined and discussed in details in the forthcoming sections. \\ 

\noindent
The paper is organized as follows: the next section is devoted to the description of the IPMC and chemico-physical process of activation is briefly described. This section reports also the way of modeling the EAP, especially, the conservation laws and constitutive equations are summarized. Section 3 focuses on the application of the model to a slender beam made of thin layer of EAP. A part of the section highlights the influence of the cation concentration within the thickness on the dielectric permittivity. Three kinds of dielectric laws will be considered. Numerical simulations are presented in Section 4. The profiles of the relevant variables and the scaling laws  are reported. Comparisons to the experimental data available in the literature are discussed according to the chosen law of the dielectric permittivity. 

\section{Modelling of the polymer}

\subsection{Description of the material and hypothesis}
\begin{figure} [h]
\includegraphics [width=0.45\textwidth]{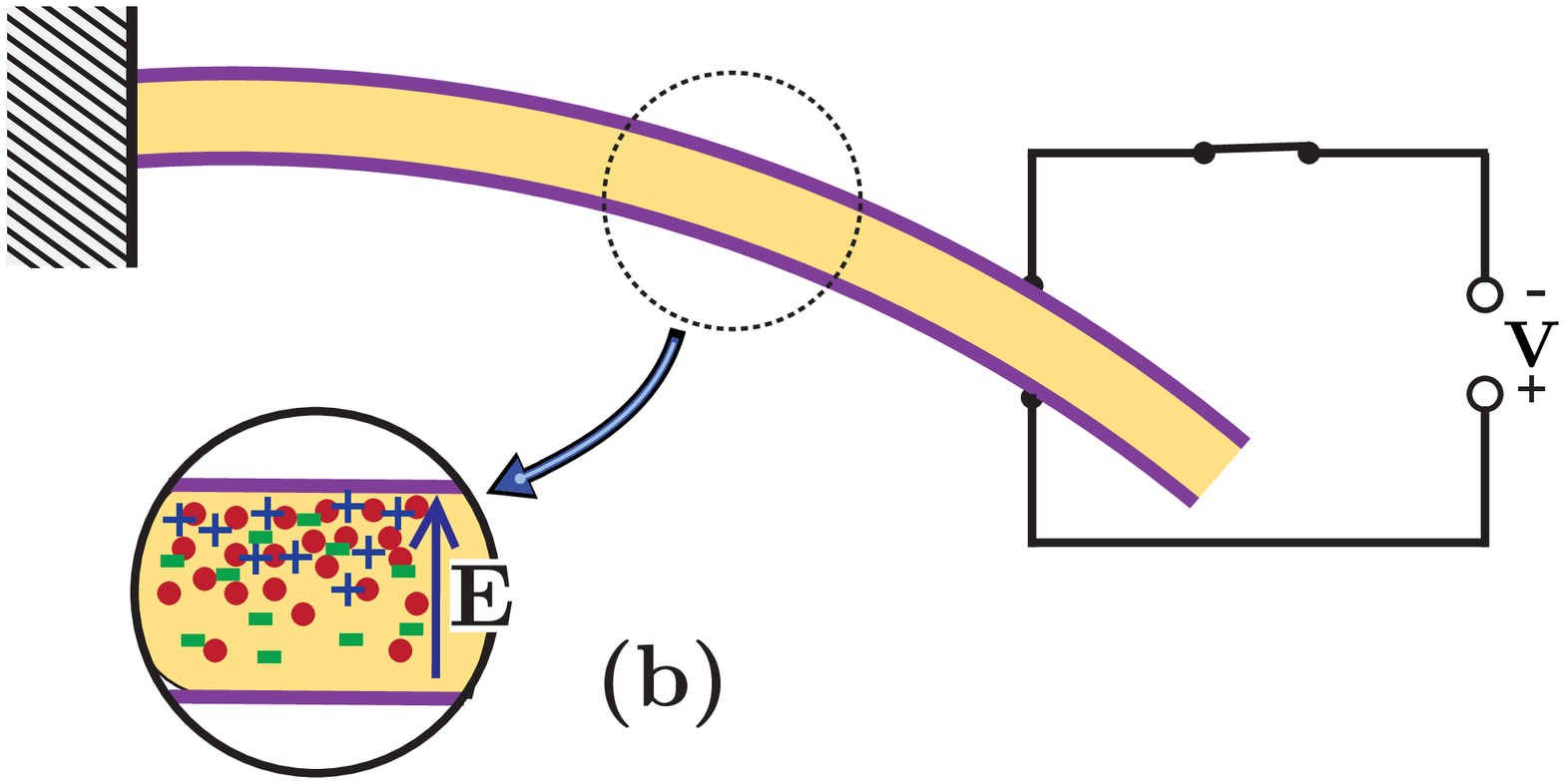}
\caption{Deformable porous medium: (a) Undeformed strip (b) Strip bending under an applied electric field }
\label{fig:1}
\end{figure}

We study an IPMC (ionic polymer-metal composite): the system consists in an ionic electro-active polymer (EAP) blade of which the two faces are covered with thin metal layers acting as electrodes. The model we developed applies for example to Nafion$^{\mbox{\scriptsize{\textregistered}}}$, an ionic polymer well documented in the literature that we will use for the validation of our constitutive equations. When saturated with water, this polymer dissociates quasi completely, releasing small cations in water whereas anions remain bound to the polymer backbone (Chabe 2008). When a potential difference is applied between the two electrodes, the cations migrate towards the negative electrode, carrying the water away by osmosis. As a result, the blade contracts on the side of the positive electrode and swells on the opposite side, causing its bending (\Fref{fig:1}).

\begin{figure} [h]
\includegraphics [width=0.45\textwidth]{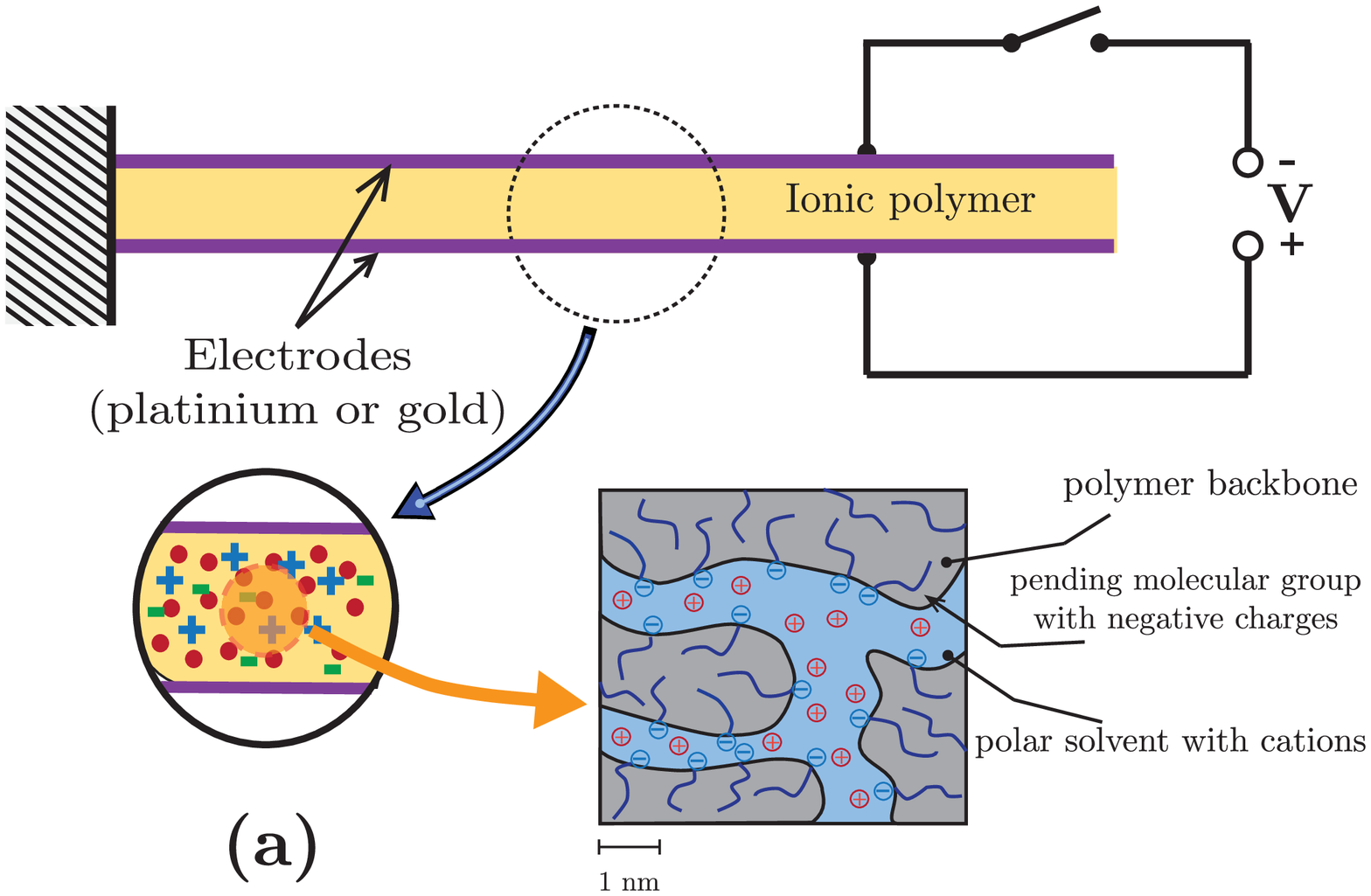}
\end{figure}

To model the electro-active polymer, we used a "continuous medium" approach. Negatively charged polymer chains are assimilated to a deformable, homogeneous and isotropic porous medium in which flows an ionic solution (water and cations). The system is therefore composed of two phases and three components which move relative to each other: the cations, the solvent (water) and the porous solid. The solid and liquid phases (water + cations) are separated by an interface without thickness. The three components are respectively identified by the subscripts $1$, $2$ and $3$; $4$ denotes the solution ($1+2$) and $i$ the interfaces.  Gravity and magnetic induction are supposed to be negligible; the only external force exerted is therefore the electric action. The different phases are supposed to be incompressible and the solution diluted. It is further recognized that solid deformations are small.

\subsection{Basis of the model}
We used a coarse-grained model developed for two-component mixtures (Ishii \& Hibiki 2006). We define two scales. The conservation equations are first written at the microscopic level for each phase and for the interfaces. At this scale (typically about $100 \; A^\circ$), the elementary volume contains one phase only but is large enough so that the medium can be considered as continuous. The macroscopic equations of the material are deduced by averaging the microscopic ones using a presence function for each phase and interface. For each phase k and for the interfaces, a Heaviside-like function of presence is defined. The macroscale quantities are obtained by averaging the corresponding microscale quantities weighted by the functions of presence. This volume average is assumed to be equivalent to a statistical average (ergodic hypothesis). At the macroscopic scale, the representative elementary volume (R.E.V.) must be large enough so that these averages are relevant, but small enough so that the average quantities could be considered as local. According to Gierke et al. (1981) and Chabe (2008), its characteristic length is about $1 \; \mu m$.

To write the balance equations, it is necessary to calculate the variations of the extensive quantities for a closed system in the thermodynamic sense. At the microscopic scale, we use the particle derivative or derivative following the motion of a constituent or an interface. At the macroscopic scale, the three constituents velocities are different; we introduce a "material derivative" or derivative following the movement of the three constituants $\frac{D}{Dt}$, which is a weighted average of the particle derivatives related to each constituent (Coussy 1995, Biot 1977).

\subsection{Conservation laws}
We thus obtain balance equations of mass \eref{m}, momentum \eref{CQ}, total, kinetic, potential and internal energy densities \eref{U}, entropy, electric charge \eref{CC} and the Maxwell equations \eref{maxwell} for the complete material (Tixier \& Pouget 2014)

\begin{equation}
\frac{\partial \rho }{\partial t}+ \mathop{\rm div} \left( \rho \overrightarrow{V}\right)=0 \,, \label{m}
\end{equation}%
\begin{equation}
\rho \frac{D\overrightarrow{V}}{Dt}= \mathop{\rm div} \utilde{\sigma }+\rho Z\overrightarrow{E}\,,  \label{CQ}
\end{equation}%
\begin{eqnarray}
\fl \rho \frac{D}{Dt}\left( \frac{U}{\rho }\right) =\sum\limits_{3,4}\left(\utilde{\sigma_{k}}:
\mathop{\rm grad} \vec{V_{k}}\right) \nonumber\\
\qquad \qquad + \left( \vec{I}-\sum\limits_{k=3,4,i}\left( \rho_{k}Z_{k}\vec{V_{k}}\right) \right)
\cdot \vec{E}- \mathop{\rm div} \vec{Q} \,, \label{U}
\end{eqnarray}
\begin{equation}
\mathop{\rm div} \overrightarrow{I}+\frac{\partial \rho Z}{\partial t}=0 \,, \label{CC}
\end{equation}%
\begin{equation}
\mathop{\rm rot} \overrightarrow{E}=\overrightarrow{0} \,, \qquad \mathop{\rm div}\overrightarrow{D}=\rho Z \,, \qquad \overrightarrow{D}%
=\varepsilon \overrightarrow{E} \,,  \label{maxwell}
\end{equation}
where $\rho_{k}$ denotes the densities relative to the volume of the whole material, $\overrightarrow{V}$ the velocity, $\utilde{\sigma}$ the stress tensor, $Z$ the electrical charge per unit of mass, $\overrightarrow{E}$ the electric field, $U$ the internal energy density, $\overrightarrow{I}$ the current density vector, $\overrightarrow{Q}$ the heat flux, $\overrightarrow{D}$ the electrical induction and $\varepsilon$ the dielectric permittivity. Subscripts refer to a phase, interface or a constituent and quantities without subscript are relative to the whole material.

We verify that the stress tensor of the whole material is symmetrical. The second member of equation \eref{U} highlights the source terms of the internal energy (viscous dissipation and Joule heating) and its flux $\overrightarrow{Q}$. Equations \eref{CC} and \eref{maxwell} show that an EAP behaves like an isotropic homogeneous linear dielectric. In the last Maxwell equation, the permittivity of the whole material is obtained by a local mixing law, and is therefore likely to vary over space and time.

\subsection{Constitutive equations}
We make the hypothesis of local thermodynamic equilibrium. The Gibbs relation of the whole material is deduced from the Gibbs relations introduced by de Groot et Mazur (1962) for a deformable solid and for a two-constituent fluid
\begin{eqnarray}
\fl T\frac{d}{dt}\left( \frac{S}{\rho }\right) =\frac{d}{dt}\left( \frac{U}{\rho }\right) +p\frac{d}{dt}\left( \frac{1}{\rho }\right) \nonumber\\
\qquad -\sum\limits_{k=1,2,3}\mu _{k}\frac{d}{dt}\left( \frac{\rho _{k}}{\rho }\right) -\frac{1}{\rho } \left( p \utilde{1} + \utilde{\sigma^{e}} \right) : \mathop{\rm grad} \overrightarrow{V} \,, \label{Gibbs}
\end{eqnarray}
where $T$ is the absolute temperature, $S$ the entropy density, $p$ the pressure, $\mu _{k}$ the mass chemical potentials, $\utilde{1} $ the identity tensor and $\utilde{\sigma ^{e}}$ the equilibrium stress tensor. $\frac{d}{dt}$ denotes the derivative following the barycentric velocity $\overrightarrow{V}$.
At equilibrium, it is assumed that the material satisfies Hooke's law and that the liquid phase is newtonian and stokesian.
 
By combining the balance equations of internal energy and entropy with the Gibbs relation, we deduced the entropy production of the system. The thermodynamics of  linear irreversible processes makes then possible to identify the generalized forces and fluxes and to deduce the constitutive equations of the electro-active polymer. According to the Curie symmetry principle, a coupling between a force and a flux of different tensorial ranks is impossible because of the isotropy of the medium.

We thus obtained a  Kelvin-Voigt type stress-strain relation \eref{rheo} and generalized Fourier's, Darcy's \eref{Darcy} and Fick's laws \eref{Nernst}. Given the orders of magnitude of the different physico-chemical parameters of the polymer (in particular, we admit that the solution is diluted), these equations can be written in the isothermal case on a first approximation (Tixier \& Pouget 2016)
\begin{equation}
\utilde{\sigma }=\lambda \left( tr\utilde{\epsilon }\right) \utilde{1}+2G%
\utilde{\epsilon }+\lambda_{v}\left( tr \dot{\utilde{\epsilon }} \right)
 \utilde{1}+2\mu_{v}\dot{\utilde{\epsilon }} \,, \label{rheo}
\end{equation}
\begin{equation}
\overrightarrow{V_{4}}-\overrightarrow{V_{3}}\simeq -\frac{K}{\eta_{2} \phi_{4}}\left[ \mathop{\rm grad} p -\left(C F - \rho _{2}^{0} Z_{3}\right) \overrightarrow{E}\right] \,, \label{Darcy}
\end{equation}%
\begin{eqnarray}
\fl \overrightarrow{V_{1}}=\overrightarrow{V_{2}} \label{Nernst}\\
-\frac{\mathsf{D}}{C}\left[\mathop{\rm grad} C-\frac{Z_{1}M_{1}C}{RT}\overrightarrow{E}+\frac{Cv_{1}}{RT}\left( 1-\frac{M_{1}}{M_{2}}\frac{v_{2}}{v_{1}}\right) \mathop{\rm grad} p\right] \,, \nonumber
\end{eqnarray}
where $\utilde{\epsilon}$ denotes the strain tensor, $\lambda$ the first Lamé constant, $G$ the shear modulus, $\lambda_{v}$ and $\mu_{v}$ the viscoelastic coefficients; $\eta_{2}$ is the dynamic viscosity of the solvent, $\phi_{k}$ the volume fractions, $K$ the intrinsic permeability of the solid, $C$ the cations molar concentration, $F = 96487 ~ C ~ mol ^ {- 1}$ the Faraday's constant, $\rho_{2}^{0}$ the mass density of the solvent, $\mathsf{D}$ the mass diffusion coefficient of the cations in the liquid phase,  $M_ {k}$ the molar masses, $v_ {k}$ the partial molar volumes and $ R = 8.31 J ~ K ^ {- 1} $ the gaz constant.

The generalized Darcy's law models the motion of the solution compared with the solid phase. This movement is caused by the pressure forces, and also by the electric field, which reflects the electro-osmosis phenomenon.
The third equation expresses the motion of cations by convection ($\overrightarrow{V_{2}}$), by mass diffusion, under the actions of the electric field and the pressure field; it can be identified with a Nernst-Planck equation (Lakshminarayanaiah 1969). An estimation of these different terms in the case of Nafion$^{\mbox{\scriptsize{\textregistered}}}Li^{+}$ shows that the pressure one is negligible.

\section{Application of the model to a static cantilevered beam}

\subsection{Beam model on large displacements}

\begin{figure} [h]
\includegraphics [width=0.45\textwidth]{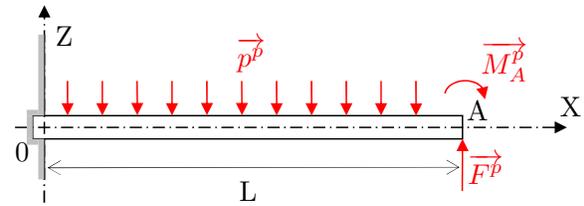}
\caption{Forces exerted on the beam}
\label{fig:2}
\end{figure}

To validate this model, we apply it to an EAP strip clamped at its end O under the action of a permanent electric field (static case). The other end $A$ is either free or subject to a shear force preventing its displacement (blocking force). Consider an EAP strip of length $L$, of width $2l$ and of thickness $2e$; we define a coordinate system $Oxyz$ such that the $Ox$ axis is along the length, the $Oy$ axis along its width and the $Oz$ axis parallel to the imposed electric field (see \fref{fig:2}). For all numerical applications, we choose the Nafion$^{\mbox{\scriptsize{\textregistered}}}Li^{+}$, an EAP well documented; in the nominal case, the dimensions of the strip are $L=2 \; cm$, $l=2.5 \; mm$ and $e=100 \; \mu m$ and it is subject to an electric potential difference $\varphi_{0}=1~V$. Considering these values and the exerted forces, this is a two-dimensional $(x,z)$ problem. The strip being thin and given the high values of the deflection, we used a beam model in large displacements.

The force system applied to the beam can be modelled by a distributed lineic force $\overrightarrow{p^{p}}$, a bending moment $\overrightarrow{M_{A}^{p}}$ applied to the end $A$ and in some cases a shear force $\overrightarrow{F^{p}}$ in $A$. The strip being very slender, we venture the hypothesis that the distributed force is independent of the $x$ coordinate along the beam and is orthogonal to it. The internal electrostatic forces of the strip cancel each other. The electric force produced by the electrodes also vanishes due to the electroneutrality condition. We deduce that the distributed force is zero everywhere.

\begin{figure} [h]
\includegraphics [width=0.4\textwidth]{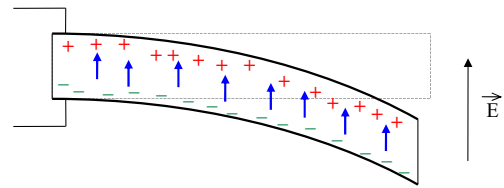}
\caption{EAP bending strip}
\label{fig:3}
\end{figure}

When a potential difference $\varphi_{0}$ is applied between the two faces, the cations and the solvent move towards the negative electrode, causing a volume variation and the bending of the strip (\fref{fig:3}). The bending moment is therefore exerted along the $Oy$ axis and results from the pressure forces $p = -\frac{\sigma_{xx}} {3}$
\begin{equation}
M_{A}^{p}= -\int_{-l}^{l} \int_{-e}^{e}\sigma_{xx}~z~dz~dy = 6l \int_{-e}^{e} p~z~dz \,.
\end{equation}

\begin{figure} [h]
\includegraphics [width=0.3\textwidth]{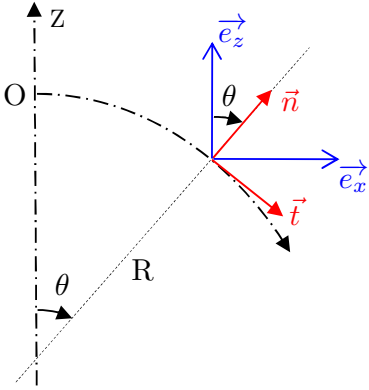}
\caption{Beam on large displacements: coordinate system}
\label{fig:4}
\end{figure}

We made the usual hypotheses of a beam model: we assumed that the straight sections of the strip remain flat and normal to the neutral fibers after deformation (Bernoulli hypothesis) and that the stress and strain distributions are independent of the application points of the external forces (Barré Saint Venant hypothesis). We define a local coordinate system where $\overrightarrow{t}$ and $\overrightarrow{n}$ are the tangent and normal vectors; $s$ and $\overline{s}$ denote the curvilinear abscissas along the beam at the rest and deformed configurations respectively, $\overline{n}$ the coordinate in the normal direction and $\theta$ the angle of rotation of a cross-section (\fref{fig:4}). Let choose the point $O$ as the origin of the curvilinear abscissas. No normal effort is applied, so we assumed that there is no beam elongation ($d \overline{s}=ds$). 

The bending moment in the current section $M^{p}$ and the radius of curvature $R^{p}$ are
\begin{equation}
M^{p} = F^{p} \left( \overline{s} -L \right) + M_{A}^{p} \,,
\qquad \qquad \frac{1}{R^{p}} = \frac{d \theta}{d \overline{s}} \,.
\end{equation}
Let $\overrightarrow{u}$ the displacement vector; its gradient with respect to the reference configuration (beam at rest) is
\begin{equation}
\mathop{\rm Grad} \overrightarrow{u} = 
\left( \begin{tabular}{cc}
$\left( 1 - \frac{\overline{n}}{R^{p}} \right) 
 \cos \theta - 1$ & $-\sin \theta$ \\
   $\left( 1 - \frac{\overline{n}}{R^{p}} \right) \sin \theta$ & 
 $\cos \theta - 1 $
\end{tabular}
\right) \,.
\end{equation}
One deduces the strain tensor
\begin{equation}
\utilde{\epsilon} = \frac{1}{2} \left[ \left( \mathop{\rm Grad} 
\overrightarrow{u} + \utilde{1} \right) ^{T} \left( 
\mathop{\rm Grad} \overrightarrow{u} + \utilde{1} \right) 
- \utilde{1} \right] \,.
\end{equation}
The beam is thin, so $\left| \overline{n} \right| << R^{p}$ and
\begin{equation}
\epsilon_{xx} = - \frac{\overline{n}}{R^{p}} \left( 1 - \frac{\overline{n}}{2R^{p}} \right) \simeq - 
\frac{\overline{n}}{R^{p}} \,.
\end{equation}
In the case of a pure bending beam, the strain is $\epsilon_{xx} = \frac {M^{p}} {EI^{p}} \overline{n} $ using the constitutive law for the bending moment $M^{p}=\frac{EI^{p}}{R^p}$, 
where $E$ is the Young's modulus and $I^{p} = \frac {4le^{3}} {3}$ the moment of inertia with respect to the $Oy$ axis. The shear force has a negligible effect on the deflection, so we deduce

\begin{equation}
\frac{1}{R^{p}} = \frac{d \theta}{d \overline{s}} = \frac{F^{p}}{E I^{p}}  \left( L-\overline{s} 
\right) - \frac{M_{A}^{p}}{E I^{p}} \,,
\end{equation}
\begin{equation}
\theta = \frac{F^{p}}{2E I^{p}} \overline{s} \left( 2L-\overline{s} \right) - \frac{M_{A}^{p}}{E I^{p}} \overline{s} \,.
\end{equation}
The deflection $w$ is obtained by integrating the relation $\frac{dz}{d \overline{s}} = \sin \theta$.

If $F^{p}=0$, the beam is circle shaped ($R^{p}$ is constant)
\begin{equation}
\begin{tabular}{cc}
$w = \frac{E I^{p}}{M_{A}^{p}} \left[ \cos \left( \frac{M_{A}^{p}}{E I^{p}} L \right) -1 \right] \,, \quad $ & $\theta = - \frac{M_{A}^{p}}{E I^{p}} L \,.$
\end{tabular}
\label{w}
\end{equation}
It should be noted that this deflection calculation becomes incorrect if the angle of rotation exceeds $90^{\circ}$. With the hypothesis of small displacements, we would have obtained
\begin{equation}
w_{s} = -\frac{M_{A}^{p}}{2 E I^{p}} L^{2} \,.
\end{equation}
If a blocking force is exerted, the deflection is zero which provides on small displacements
\begin{equation}
F^{p}_{s} = \frac{3 M_{A}^{p}}{2L} \,.
\label{FP}
\end{equation}
On large displacements, $F^{p}$ verify
\begin{equation}
w = \int_{0}^{L} \sin \left[\frac{F^{p}}{2E I^{p}} x (x-2L) + \frac{M_{A}^{p}}{E I^{p}} x \right]~dx = 0 \,.
\label{Fg}
\end{equation}

Let us compare this result with the calculation on small displacements. The above condition can be written
\begin{equation}
\begin{tabular}{ll}
$w=0$&$=\int_{0}^{L}\sin \left[ \frac{1}{x^{\ast ^{2}}}\left(
x+x_{0}\right) ^{2}-\frac{x_{0}^{2}}{x^{\ast ^{2}}}\right] \;dx$\\
&$ =x^{\ast }\cos \left( \frac{x_{0}^{2}}{x^{\ast ^{2}}}\right) \left[
S\left( x_{1}/x^{\ast }\right) -S\left( x_{0}/x^{\ast }\right) \right]$\\
&$ -x^{\ast }\sin \left( \frac{x_{0}^{2}}{x^{\ast ^{2}}}\right) \left[ C\left(x_{1}/x^{\ast }\right) -C\left( x_{0}/x^{\ast }\right) \right] \,, $
\end{tabular}
\end{equation}
where 
\begin{equation}
\begin{tabular}{lll}
$x^{\ast}=\sqrt{\frac{2EI^{p}}{F^{p}}} \,, \quad $ & $x_{0}=\frac{%
M_{A}^{p}}{F^{p}}-L \,, \quad $ & $x_{1}=\frac{M_{A}^{p}}{F^{p}} \,,$%
\end{tabular}%
\end{equation}
and where $ S $ and $ C $ denote the Fresnel functions
\begin{equation}
\begin{tabular}{ll}
$S(x)=\int_{0}^{x}\sin t^{2}\;dt$ & $=\sum\limits_{n=0}^{+\infty
}\left( -1\right) ^{n}\frac{x^{4n+3}}{\left( 2n+1\right) ! \; \left( 4n+3\right) 
} \,, $ \\ 
$C(x)=\int_{0}^{x}\cos s^{2}\;dt$ & $=\sum\limits_{n=0}^{+\infty
}\left( -1\right) ^{n}\frac{x^{4n+1}}{\left( 2n\right) ! \; \left( 4n+1\right) } \,. $%
\end{tabular}%
\end{equation}
In the case of Nafion$^{\mbox{\scriptsize{\textregistered}}}Li^{+}$, $E = 1,3 \; 10^{8} \; Pa$ (Bauer et al. 2005, Barclay Satterfield \& Benziger 2009, Silberstein \& Boyce 2010). Using the blocking force values provided by the literature in the nominal case (Newbury 2002, Newbury \& Leo 2002, Newbury \& Leo 2003), $\frac{x_{0}}{x^{\ast }}<\frac{x_{1}}{x^{\ast }} \sim 0.45$. We deduce, with a relative error around $2 \%$
\begin{equation}
w \simeq \frac{F^{p}L^{2}}{6EI^{p}}\left( 3\frac{M_{A}^{p}}{F^{p}}%
-2L\right) \simeq 0 \,.
\end{equation}
As a consequence, small displacements give a good approximation of the blocking force, that we will verify with our numerical simulations. On the contrary, the calculation of the deflection of the cantilevered beam requires a large displacements model.

\subsection{Additional hypothesis and static equations for a bending strip}
Let us evaluate the variations of the volume fraction of the solution $\phi_{4}$. Consider a small volume $dV$ located at a distance $z$ from the beam axis. According to Bernoulli's hypothesis, this volume takes the value $\frac{\left| R^{p} \right|+z}{\left| R^{p} \right|} dV$ when the beam bends with a radius of curvature $R^{p}$. The volume of the solid phase does not change, and the variation of liquid phase volume fraction is about
\begin{equation}
d\phi_{4}=\frac {\phi_{3} z} {\left| R^{p} \right|}
=(1-\phi_{4}) \frac{M^{p}}{EI^{p}} z
\simeq (1-\phi_{4}) \frac{2 \vert w \vert}{L^{2}} z \,.
\end{equation}
$\phi_{4} \simeq 0.38$ (Cappadonia et al. 1994, Chabé 2008, Nemat-Nasser \& Li 2000). In the nominal case, $w \sim 1 \; mm$ (Nemat-Nasser 2002, Newbury \& Leo 2002, Newbury 2002). $\phi_{4}$ therefore varies less than $0.1\%$ over the thickness of the beam. As a consequence, we assume that the volume fraction $\phi_{4}$ is constant.

Considering the dimensions of the strip, we assume that the problem is two-dimensional in the $Oxz$ plane. On a first approximation, we suppose that the electric field and induction are parallel to the $Oz$ axis: $\overrightarrow{E} \simeq E_ {z} \overrightarrow{e_{z}} $ and $\overrightarrow{D} \simeq D_{z} \overrightarrow {e_{z}} $. We further admit that $C$, $E_{z}$, $D_{z}$, $p$, $\rho Z$ and the electrical potential $\varphi$ only depend on the variable $z$. Finally, we neglect the pressure term of the equation (\ref{Nernst}), an assumption that we will verify later. The equation system of our model then becomes
\begin{equation}
\begin{tabular}{ll}
$E_{z}= - \frac {d \varphi} {dz} \,, $ & $\frac {d D_{z}} {dz} = \rho Z  \,,$ \smallskip \\
$D_{z} = \varepsilon E_{z} \,,$ & $\rho Z = \phi_{4} F \left(C - C_{moy} \right) \,, $ \smallskip \\
$\frac {dp} {dz} = \left(C F - \rho _{2}^{0} Z_{3}\right) E_{z} \,,$ & $\frac {dC} {dz} = \frac{FC}{RT} E_{z} \,,$ \\
\end{tabular} \label{Eqp}
\end{equation}
where
\begin{equation}
C_{moy}=-\frac{\left( 1-\phi _{4}\right) \rho _{3}^{0}Z_{3}}{\phi _{4}F} \,.
\end{equation}
$C_{moy}$ denotes the cations average concentration. The anions being attached to the polymer chains, they are uniformly distributed within the material; their concentration is therefore constant and equal to $C_{moy}$ considering the electroneutrality condition. From Nemat-Nasser \& Li (2000), the mass density of dry Nafion is $\rho_{3}^{0}=2 \; 10^{3} \; kg \; m^{-3}$, and its equivalent weight, that is the weight of polymer per mole of sulfonate groups, is $M_{eq}=1
.1 \; kg \; eq^{-1}$ (Chabé 2008, Colette 2008), which provides $Z_{3}=-\frac{F}{M_{eq}}=-9 \; 10^{-4} \; C \; kg^{-1}$. We deduce $C_{moy}=3080 \; mol \; m^{-3}$. We also choose an absolute temperature $T=300~K$.

The boundary conditions and the electroneutrality condition are
\begin{equation}
\begin{tabular}{ccc}
$\varphi(-e) =\varphi_{0} \,, \quad $ & $%
\varphi (e) = 0 \,, \quad $ & $
\int_{-e}^{e} \rho Z \, \mathrm{d}z =0 \,.$
\end{tabular} \label{CL}
\end{equation}
According to (\ref{Eqp}), this last condition is equivalent to $D_{z}\left( e \right) = D_{z}\left( -e \right)$.

The permittivity strongly depends on the conductivity, hence of the electric charge; it increases with the water uptake (Deng \& Mauritz 1992, Nemat-Nasser 2002), therefore with the cations concentration. We assume that it satisfies a mixing law
\begin{equation}
\varepsilon = \varepsilon^{0} + \alpha C \,,
\qquad with \qquad
\alpha = \frac{\varepsilon_{moy}-\varepsilon^{0}}{C_{moy}} \,,
\end{equation}
where $\varepsilon_{moy}$ denotes the average permittivity of the material.
We have considered three models of permittivity: constant, linear and affine. We choose our permittivity values in order that deflections and blocking forces are in agreement with the literature data, namely $0.5<w<1.5 \; mm$ (Nemat-Nasser 2002, Newbury 2002) and $0.6<F^{p}<1.3 \; mN$ (Newbury 2002, Newbury \& Leo 2002, Newbury \& Leo 2003) in the nominal case (\tref{t1}).

\begin{table}
\caption{\label{t1} Dielectric permittivity values}
\begin{indented}
\item[]\begin{tabular}{@{}lll}
\br
&$\varepsilon^{0}~(Fm^{-1})$&$\varepsilon_{moy}~(Fm^{-1})$\\
\mr
constant & $5~10^{-7}$&$5 ~10^{-7}$\\
linear & $0$&$10^{-4}$\\
affine & $5 \;10^{-7}$&$10^{-4}$\\
\br
\end{tabular}
\end{indented}
\end{table}

The Nafion relative permittivity found in the literature are very scattered, but compatible with those we have chosen. The average permittivity was measured by Deng \& Mauritz (1992) for hydrated perfluorosulfonate ionomer membranes of the Nafion$^{\mbox{\scriptsize{\textregistered}}}$ family with different water contents. They obtained permittivity values between $10^{-7}~F~m^{- 1}$ and $10^{-6}~F~m^{- 1}$ by electrical impedance measurements. Nemat-Nasser (2002) deduced the permittivity of capacity measurements by assimilating the IPMC strip to a capacitor; for Nafion$^{\mbox{\scriptsize{\textregistered}}}$ $117~Li^{+}$, he got $\varepsilon \simeq 2.7~10^{- 3}~F ~ m^{- 1}$. Farinholt \& Leo (2014) deduced a close value from their measurements but did not specify their method. Wang et al. (2014) measured the permittivity of Nafion$^{\mbox{\scriptsize{\textregistered}}}$ $117~Na^{+}$ by time domain dielectric spectroscopy. For samples obtained with different manufacturing methods and a  water content about $22\%$, the permittivity ranges from $5~10^{-5}$ to $5~10^{-3}$.

\subsection{Resolution with different permittivity models}
Let us introduce dimensionless variables
\begin{equation}
\begin{tabular}{lll}
$\overline{E} = \frac {E_{z} e} {\varphi_{0}} \,,$ & $ \overline{\varphi} = \frac {\varphi} {\varphi_{0}} \,,$ &$\overline{D}=\frac{D_{z}}{e \phi_{4} FC_{moy}} \,,$ \smallskip\\
$ \overline{C} = \frac {C} {C_{moy}} \,,$&$\overline{\rho Z} = \frac {\rho Z} {\phi_{4} F C_{moy}} \,,$&$\overline{p} = \frac {p} {F \varphi_{0} C_{moy}} \,, \smallskip$\\
$ \overline{z} = \frac {z} {e} \,,$&$\overline{\varepsilon}=\frac{\varepsilon \varphi_{0}}{e^{2} \phi_{4} FC_{moy}} \,.$ \\
\end{tabular}
\end{equation}
The system of equations and the boundary conditions become
\begin{equation}
\begin{tabular}{ll}
$\overline{E}=-\frac{d\overline{\varphi }}{d\overline{z}} \,, \quad $& $\overline{D} = \overline{\varepsilon} \overline{E} \,,$ \smallskip\\ 
$\frac{d\overline{D}}{d\overline{z}}=\overline{\rho Z} \,, \quad $ &$\overline{\varepsilon} = A_{0} \overline{C} + A_{1} \,,$ \smallskip\\
$\frac{d\overline{p}}{d\overline{z}}=\left( \overline{C}+A_{3}\right) \overline{E} \,, \quad $& $\frac{d\overline{C}}{d\overline{z}}=A_{2} \overline{C} \overline{E} \,,$ \smallskip\\
$\overline{\rho Z}=\overline{C}-1 \,,$ & $\overline{D}(1)=\overline{D}(-1) \,,$ \smallskip\\
$ \overline{\varphi }(-1)=1 \,,$&$\overline{\varphi }(1)=0 \,,$\\
\end{tabular}
\label{syst}
\end{equation}
with
\begin{equation}
\begin{tabular}{ll}
$A_{0}=\frac{\varphi_{0} (\varepsilon_{moy}-\varepsilon^{0})}{e^{2}\phi_{4}FC_{moy}} \,,\quad$ & $A_{1}=\frac{\varphi_{0} \varepsilon^{0}}{e^{2}\phi_{4}FC_{moy}} \,, \smallskip$ \\
$A_{2}=\frac{F\varphi _{0}}{RT}\sim 38.7 \,, \quad$ & $A_{3}=-\frac{\rho _{2}^{0}Z_{3}}{C_{moy}F}\sim 0.303 \,,$ \\ 
\end{tabular}%
\end{equation}
in the nominal case. We deduce the following relations
\begin{equation}
\overline{C}=B_{1}\exp \left( -A_{2}\overline{\varphi }\right) \,,
\end{equation}
\begin{equation}
\frac{\overline{D}^{2}}{2} = \frac{1}{A_{2}} \left( \frac{A_{0}}{2} \overline{C}^{2}+(A_{1}-A_{0}) \overline{C} \right) + A_{1} \overline{\varphi} + \frac{B_{2}}{2} \,,
\end{equation}
\begin{equation}
\overline{p}=\frac{\overline{C}}{A_{2}}-A_{3}\overline{\varphi }+B_{3} \,,
\end{equation}
where $B_{1}$, $B_{2}$ and $B_{3}$ are three constants.
The polymer strip behaves like a conductive material. The electric field, displacement and charge are then zero throughout the strip except near the sides. We can deduce
\begin{equation}
B_{2}=\frac{1}{A_{2}} (A_{0}-2 A_{1} - 2 A_{1} \ln B_{1}) \,.
\end{equation}

The positive constant $B_{1}$ satisfies the electroneutrality condition (\ref{syst})
\begin{equation}
A_{0} (1+ e^{-A_{2}})B_{1}^{2} +2(A_{1}-A_{0}) B_{1} -\frac{2 A_{1}A_{2}}{1- e^{-A_{2}}}= 0 \,.
\label{B1comp}
\end{equation}
When $\varphi_{0} \gtrsim 1 \; V$, $e^{-A_{2}}<<1$. In the case of a constant permittivity ($A_{0}=0$)
\begin{equation}
B_{1} = \frac{ A_{2}}{1-e^{-A_{2}}} \simeq A_{2} \,.
\end{equation}
In case of a linear permittivity ($A_{1}=0$)
\begin{equation}
B_{1} =  \frac{2}{1+e^{-A_{2}}} \simeq 2 \,,
\end{equation}
and in the general case of an affine permittivity $B_{1}$ is of the order of $2$
\begin{equation}
\begin{tabular}{ll}
$B_{1} $&$= \frac{A_{0}-A_{1}}{A_{0}(1+e^{-A_{2}})} \left[ 1+ \sqrt{1+ \frac{2 A_{0} A_{1} A_{2}}{(A_{0}-A_{1})^{2}}\frac{1+e^{-A_{2}}}{1-e^{-A_{2}}}} \right]$ \smallskip \\
&$\sim 2 \,,$
\end{tabular}
\end{equation}
assuming $A_{1}<<A_{0}$ and $\frac{2 A_{2} A_{1}}{A_{0}} <<1$, assumption checked since we choose $\varepsilon^{0}<<\varepsilon_{moy}$. We moreover verify that $B_{1}$ tends toward $1$ when $\varphi_{0}$ tends toward $0$.

The system of equations (\ref{syst}) enable us to calculate the values of the various parameters at the center and at the boundaries of the strip (\tref{t2}). Note that $\overline{p}$ is defined to within an additive constant.

\begin{table}
\caption{\label{t2} Center and boundary values of the different quantities}
\begin{indented}
\item[]\begin{tabular}{@{}llll}
\br
& $-1$ & center & $1$ \\
\mr
$\overline{C}$ & $B_{1}\exp \left( -A_{2}\right)$ & $1$ & $B_{1}$ \\
$\overline{\rho Z}$ & $B_{1}\exp \left( -A_{2}\right)-1$ & $0$ & $B_{1}-1$ \\
$\overline{\varphi }$ & $1$ & $\frac{\ln B_{1}}{A_{2}}$ & $0$ \\
$\overline{D}$ & $\overline{D}(1)$ & $0$ & $\overline{D}(1)$ \\
$\overline{\varepsilon}$ & $A_{1} $ & $A_{0}+A_{1}$ & $A_{0} B_{1} +A_{1} $ \\
$\overline{p}-B_{3}$ & $\frac{B_{1}\exp \left( -A_{2}\right) }{A_{2}}-A_{3}$ & $\frac{1}{A_{2}}-\frac{A_{3}\ln B_{1}}{A_{2}}$ & $\frac{B_{1}}{A_{2}}$\\
\br
\end{tabular}
where $\overline{D}(1)=\sqrt{ \frac{1}{A_{2}} \left[  A_{0} + 2 A_{1} (A_{2}-1- \ln B_{1}) \right]}$
\end{indented}
\end{table}

The bending moment is given by
\begin{equation}
\begin{tabular}{ll}
&$M^{p}_{A}=A_{5} \overline{M}
= A_{5} \int_{-1}^{1} (\frac{\overline{C}}{A_{2}}-A_{3}\overline{\varphi }) \overline{z} \;d\overline{z} \,,$ \medskip\\
with &$\overline{M} = \int_{-1}^{1} \overline{p} \, \overline{z} \;d\overline{z} \,, \smallskip$\\
&$A_{5}=6le^{2}F\varphi_{0}C_{moy}\,,$ \smallskip\\
&$\int_{-1}^{1} \overline{C} \overline{z} \;d\overline{z}
= 2 \overline{D}(1) - \frac{A_{0}}{A_{2}} B_{1} - A_{1} \,.$\\
\end{tabular}
\label{Mp}
\end{equation}
$I_{\varphi}=\int_{-1}^{1} \overline{\varphi} \, \overline{z} \;d\overline{z}$ must be numerically evaluated.
We deduce $w$, $w_{s}$, $F^{p}$, $F^{p}_{s}$ and $\theta$ using (\ref{w}) to (\ref{Fg}).

The resolution of the equation system is tricky from a numerical point of view because of the steepness of the functions near the boundaries. We used different methods according to the permittivity model to obtain the best precision.

In the case of a constant permittivity ($A_{0}=0$), we use the variable $y = \ln \overline{C}$ which verifies
\begin{equation}
\begin{tabular}{l}
$\frac{d^{2}y}{d\overline{z}^{2}} = \frac{A_{2}}{A_{1}}(e^{y}-1) \,,$\\
$\left. \frac{dy}{d\overline{z}} \right|_{1} \simeq \left. \frac{dy}{d\overline{z}} \right|_{-1} =\sqrt{2 \frac{A_{2}}{A_{1}}(A_{2}-ln A_{2}-1}) \,, $\\
$I_{\varphi}=\int_{-1}^{1} \overline{\varphi} \, \overline{z} \;d\overline{z} =-\frac{1}{A_{2}} \int_{-1}^{1} y \overline{z} \;d\overline{z} \,.$
\end{tabular}
\label{y}
\end{equation}
This differential equation can be numerically integrated near the boundaries, but not over the entire thickness.

In the case of a linear permittivity ($A_{1}=0$), the cation concentration can be calculated using the following differential equation
\begin{equation}
\frac{d^{2}\overline{C}}{d\overline{z}^{2}}=\frac{1}{\overline{z_{0}}^{2}} \left( \overline{C}-1\right) \,,
\end{equation}
where $\overline{z_{0}}=\sqrt{\frac{A_{0}}{A_{2}}}<<1$, which provides
\begin{equation}
\overline{C} = 1 + (B_{1}-1) \frac{\sinh\left( \overline{z}/\overline{z_{0}} \right)}{\sinh\left( 1/\overline{z_{0}} \right)} \,.
\end{equation}
The functions thus obtained are especially steep near the boundaries in this case, which makes the calculation of $I_{\varphi}$ very delicate; we calculate the piecewise integral using Taylor expansions on some intervals.

In the case of an affine permittivity, the following equation can be numerically integrated over the entire interval
\begin{equation}
\frac{d^{2} \overline{D}}{d \overline{z}^{2}} = \frac{A_{2}\left( 1+ \frac{d \overline{D}}{d \overline{z}} \right)}{A_{0} \left( 1+ \frac{d \overline{D}}{d \overline{z}} \right) + A_{1}} \overline{D} \,.
\end{equation}

\section{Simulation results}

\subsection{Hypothesis validation}
The dimensionless equation (\ref{Nernst}) provides
\begin{equation}
\frac{d\overline{C}}{d\overline{z}}=A_{2}\overline{C} \, \overline{E}\left[
1+A_{4}\left( \overline{C}+A_{3}\right) \right] \,,
\end{equation}
where (Tixier \& Pouget 2016)
\begin{equation*}
A_{4}=C_{moy}v_{1}\left( 1-\frac{M_{1}}{M_{2}}\frac{v_{2}}{v_{1}}\right)\simeq \frac{C_{moy}^{2} M_{1} M_{2}}{\rho_{2}^{02}} \simeq 1.2 ~10^{-3} \,,
\end{equation*}
($M_{1}=6.9~g~mol^{-1}$ and $M_{2}=18~g~mol^{-1}$ for Nafion$^{\mbox{\scriptsize{\textregistered}}}Li^{+}$). $A_{4}\left( \overline{C}+A_{3}\right) $ corresponds to the pressure term; its maximum value (about $0.04$) is reached near the negative electrode in the case of a constant permittivity, and it is of the order of $A_{4}<<1$ in the center of the strip. We can therefore neglect the pressure term for the calculation of the bending moment.

For each of our simulations, we calculated the different parameters in small and large displacements.
The difference between blocking forces calculated in small and large displacements is less than  $1 \%$ in all simulations and largely below $0.1 \%$ as soon as the thickness exceeds $50 \; \mu m$. Small displacements are thus sufficiently accurate for the blocking force calculation.
Concerning the deflection, the difference between both models is less than $1 \%$ when the deflection is less than $3.5~mm$; it otherwise becomes very large (over $100 \%$ sometimes, especially for low thicknesses). The large displacements model is thereby essential for the deflection evaluation. Moreover, we can verify that the small displacements calculation is an upper bound on deflection.

\subsection{Profiles of different quantities}

\begin{figure*} [h!]
\includegraphics[width=0.8\textwidth]{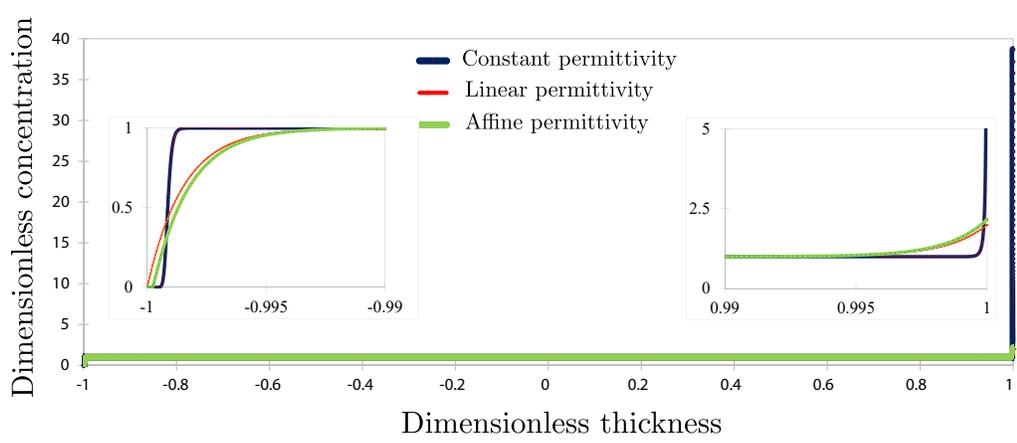}
\caption{Variation of the dimensionless cation concentration in the thickness of the strip; the distribution close to the boundaries are detailled in insets. The constant permittivity model is in blue, the linear one in red and the affine one in green.}
\label{fig:5}
\end{figure*}

\begin{figure*} [h]
\includegraphics[width=0.8\textwidth]{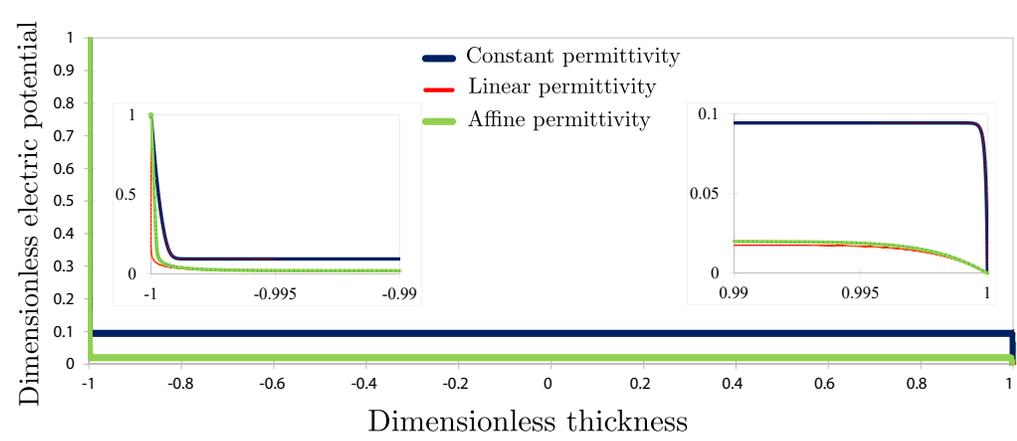}
\caption{Variation of the dimensionless electric potential in the thickness of the strip; the distribution close to the boundaries are detailled in insets (same colours as in \fref{fig:5}).}
\label{fig:6}
\end{figure*}

\begin{figure*} [h]
\includegraphics[width=0.8\textwidth]{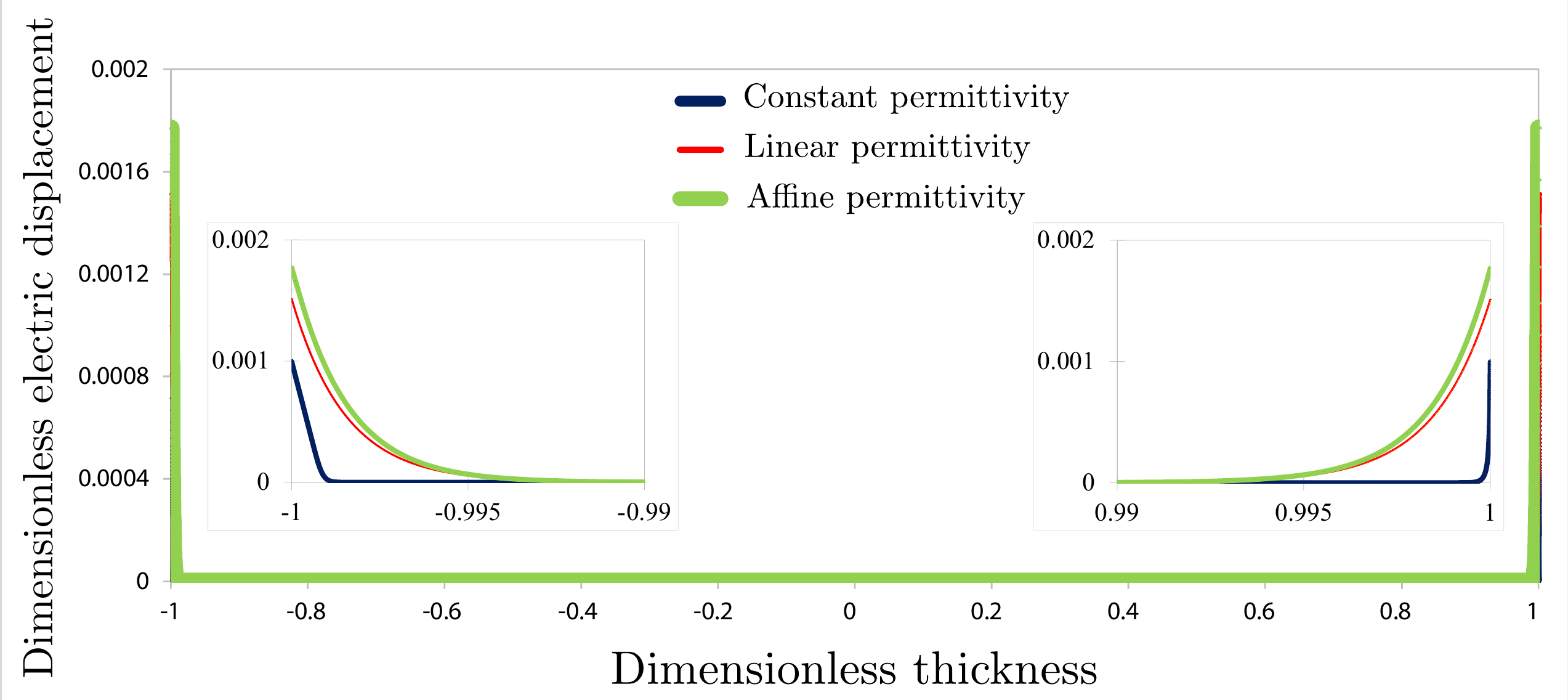}
\caption{Variation of the dimensionless electric displacement in the thickness of the strip; the distribution close to the boundaries are detailled in insets (same colours as in \fref{fig:5}).}
\label{fig:7}
\end{figure*}

\begin{figure*} [h]
\includegraphics[width=0.8\textwidth]{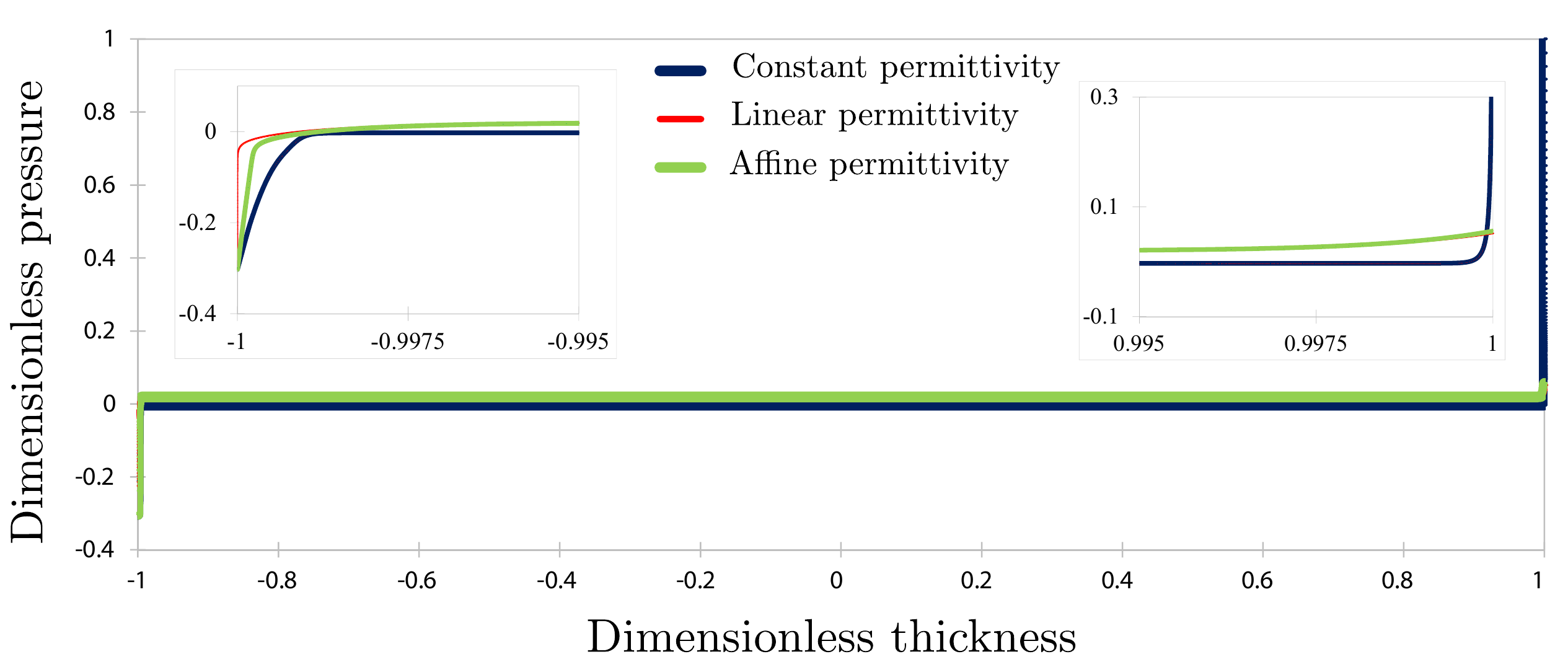}
\caption{Variation of the dimensionless pressure in the thickness of the strip; the distribution close to the boundaries are detailled in insets (same colours as in \fref{fig:5}).}
\label{fig:8}
\end{figure*}

We have plotted the profiles of the various quantities in the thickness of the strip in the nominal case described in section 3 (Nafion$^{\mbox{\scriptsize{\textregistered}}}Li^{+}$ with $L=2~cm$, $l=2.5~mm$, $e=100~\mu m$ and $\varphi_{0}=1~V$). The profiles of dimensionless concentration, electric potential and displacement and pressure have similar aspects: the curves are almost constant in the central zone and very steep near the boundaries (figures \ref{fig:5} to \ref{fig:8}).

The cation concentration profiles displays near the positive electrode a zone almost free of cations whose size depends on the permittivity model: $0.08 \; \mu m$ in the constant case and $0.03 \; \mu m$ in the affine one. The linear model predicts a permittivity tending towards $0$ and is therefore incorrect in this range.
Near the negative electrode, there is an accumulation of cations over a characteristic length depending on the chosen permittivity model: close to $0.1~\mu m$ for a constant permittivity and $1~\mu m$ in linear and affine cases. The concentration on the negative electrode is twenty times higher with the constant permittivity model than in the two other cases. Nemat-Nasser (2002), Wallmersperger et al. (2009) and Nardinocchi et al. (2011) obtained a similar profile, although less steep.

The electric potential profiles look similar to those obtained by Wallmersperger et al. (2009) and Nardinocchi et al. (2011), although they are less steep in the vicinity of the electrodes.
The linear and affine models give almost identical results for the other profiles. The constant model distinguishes by its steepness near the boundaries, with a characteristic length close to $0.02~\mu m$ near the negative electrode, twenty times smaller than the other models; near the positive electrode, its characteristic length is $0.1~\mu m$, five times larger than the other models for electric potential and pressure profiles. All three models provide nearly identical electrical displacement profiles with close boundary values.
The pressure profiles are also very similar, except near the negative electrode where the constant model provides a boundary twenty times higher than the other two.

We furthermore verify that the electrical displacement is null in the central part of the strip with the three models, which confirms that the material behaves like a conductor.

\subsection{Expected scaling laws}
Let us determine the dependence of the mechanical quantities with respect to $l$, $L$, $e$ and $\varphi_{0}$. These quantities are written
\begin{equation}
\begin{tabular}{l}
$M_{A}^{p}=6FC_{moy}~le^{2}~\varphi _{0}~\overline{M} \,,$ \smallskip\\ 
$\theta =-\frac{9}{2}\frac{FC_{moy}}{E}~\frac{L}{e}~\varphi _{0}~\overline{M} \,,$ \smallskip\\ 
$F^{p}=9FC_{moy}~\frac{le^{2}}{L}~\varphi _{0}~\overline{M} \,,$ \smallskip\\
$w_{s}=-\frac{9}{4}\frac{FC_{moy}}{E}~\frac{L^{2}}{e}~\varphi _{0}~
\overline{M} \,,$ \smallskip\\ 
$w=\frac{2}{9}\frac{E}{FC_{moy}} \frac{e}{\varphi _{0}} \frac{1}{\overline{M}}~\left[ \cos \left( \frac{9}{2} \frac{FC_{moy}}{E} \frac{L}{e} \varphi _{0} \overline{M}\right) -1\right]  \,,$ \\ 
\end{tabular}
\label{Meca}
\end{equation}%
where $F$ is a constant. $\overline{M}=\int_{-1}^{1} (\frac{\overline{C}}{A_{2}}-A_{3}\overline{\varphi })\overline{z}d\overline{z}$ is a function of the constants $A_{0}$, $A_{1}$, $A_{2}$, $A_{3}$ et $B_{1}$. Let
\begin{equation}
A_{0}=a_{0}\frac{\varphi_{0}}{e^{2}} \,, \qquad
A_{1}=a_{1}\frac{\varphi_{0}}{e^{2}} \,, \qquad
A_{2}=a_{2} \varphi_{0} \,. \qquad
\end{equation}
$a_{0}$, $a_{1}$, $a_{2}$ and $A_{3}$ like $C_{moy}$ and $E$ only depend on the chosen material.

Let us first look at the case where $\varphi_{0} \gtrsim 1$; $B_{1}$ checks the approximate equation (see eq. (\ref{B1comp}))
\begin{equation}
a_{0} B_{1}(B_{1}-2)=2 a_{1}(a_{2}\varphi_{0}-B_{1}) \,,
\end{equation}
and therefore depends only on $\varphi_ {0}$ and the material.
The pressure profile is almost constant throughout the center of the strip and is very steep near the boundaries. It can be modelled by a constant between two values $-\overline{z_{1}} $ and $\overline{z_{2}}$ with $ \overline{z_{1}}, \overline{z_{2}} \lesssim 0.99$. Assuming for example $0 <\overline{z_{1}} <\overline{z_{2}}$
\begin{equation}
\begin{tabular}{l}
$\overline{M} =\int_{-1}^{-\overline{z_{1}}}\overline{p}\;\overline{z}\;d\overline{z}+\int_{-\overline{z_{1}}}^{\overline{z_{1}}}\overline{p}\;\overline{z}\;d\overline{z}$\\
$\qquad \qquad \qquad \qquad +\int_{\overline{z_{1}}}^{\overline{z_{2}}}\overline{p}\;\overline{z}\;d\overline{z}+\int_{\overline{z_{2}}}^{1}\overline{p}\;\overline{z}\;d\overline{z} \,.$
\end{tabular}
\end{equation}
Given the low values of $\delta _{1}=1-\overline{z_{1}}$ and $\delta _{2}=1-\overline{z_{2}}$, we can expand in Taylor series up to the second order the first and last integrals
\begin{equation}
\begin{tabular}{l}
$\int_{-1}^{-\overline{z_{1}}}\overline{p}\;\overline{z}\;d\overline{z}=-\overline{p}(-1)\delta _{1}+o\left( \delta_{1}^{2}\right) \,, \smallskip$\\
$\int_{-\overline{z_{1}}}^{\overline{z_{1}}}\overline{p}\;\overline{z}\;d\overline{z}=0 \,, \smallskip$\\
$\int_{\overline{z_{1}}}^{\overline{z_{2}}}\overline{p}\;\overline{z}\;d\overline{z}=\frac{\overline{p(0)}}{2}\left( \overline{z_{2}}^{2}-\overline{z_{1}}^{2}\right) \,, \smallskip$\\
$\int_{\overline{z_{2}}}^{1}\overline{p}\;\overline{z}\;d\overline{z} = \overline{p}(1)\delta _{2}+o\left( \delta _{2}^{2}\right) \,.$
\end{tabular}
\end{equation}
Hence, in all cases and at the first order in $\delta$
\begin{equation}
\overline{M}=A_{3}\delta _{1}+\frac{B_{1}}{A_{2}} \delta _{2}+\frac{1}{A_{2}}\left( 1-A_{3}\ln B_{1}\right)
\left( \delta _{1}-\delta _{2}\right) \,.
\end{equation}
$\delta _{1}$ and $\delta _{2}$ can be roughly evaluated using the following formulas (thanks to eq. (\ref{syst}))
\begin{equation}
\left. \frac{d\overline{p}}{d\overline{z}} \right|_{-1}\sim \frac{\overline{p}\left( -\overline{z_{1}}\right) -\overline{p}(-1)}{\delta_{1}} \,, \quad \left. \frac{d\overline{p}}{d\overline{z}}\right|_{1}\sim \frac{\overline{p}(1)-\overline{p}\left( \overline{z_{2}}\right)}{\delta_{2}} \,.
\end{equation}
$\overline{M}$ is therefore independent of $L$ and $l$ and approximately inversely proportional to $e$. Its dependence on $\varphi_{0}$ is more complex and linked with the chosen permittivity model. Given the values of $a_{0}$, $a_{1}$, $a_{2}$, $A_{3}$ and $B_{1}$ in the nominal case, we obtain with a precision of about $20 \%$
\begin{equation}
\begin{tabular}{ll}
Constant case&$\overline{M} \simeq \sqrt{\frac{a_{1}}{2}} a_{3} \frac{\sqrt{\varphi_{0}}}{e}  \,,$ \smallskip \\
Linear case&$\overline{M} \simeq \sqrt{ \frac{a_{0}}{a_{2}^{3} }} \frac{1}{e \varphi_{0}} \,,$ \smallskip \\
Affine case&$\overline{M} \simeq \sqrt{ \frac{a_{0}}
{a_{2}^{3} }} \; \frac{f(\varphi_{0})}{e \varphi_{0}} \,,$ \smallskip \\
with &$f(\varphi_{0}) = \frac{\frac{a_{1}}{a_{0}} a_{2}^{2} a_{3} \varphi_{0}^{2}+2 \frac{a_{1}}{a_{0}} a_{2} \varphi_{0}+1} {\sqrt{1+2 \frac{a_{1}}{a_{0}} a_{2} \varphi_{0} }} \,.$
\end{tabular}
\label{lissage}
\end{equation}
We deduce the scaling laws presented in \tref{t3}.

\begin{table}
\caption{\label{t3} Expected scaling laws}
\begin{indented}
\item[]\begin{tabular}{@{}lllllll}
\br
& $L$ & $l$ & $e$ & $\varphi_{0}$ & $\varphi_{0}$ & $\varphi_{0}$\\
$\varepsilon$ & & & & constant & linear & affine\\
\mr
$M_{A}^{p}$ & - & $l$ & $e$ &$\varphi_{0}^{3/2}$& - &$f(\varphi_{0})$ \\ 
$\theta $ & $L$ & - & $e^{-2}$&$\varphi_{0}^{3/2}$& - &$f(\varphi_{0})$ \\ 
$F^{p}$ & $L^{-1}$ & $l$ & $e$ &$\varphi_{0}^{3/2}$& - &$f(\varphi_{0})$\\
$w_{s}$ & $L^{2}$ & - & $e^{-2}$&$\varphi_{0}^{3/2}$& - &$f(\varphi_{0})$\\
\br
\end{tabular}
\end{indented}
\end{table}

When the imposed electric potential is very small, $B_{1}$ tends to $1$ for all permittivity models, $C$ tends to $C_{moy}$ and $\varphi$ to $\varphi_{0}$ over the entire thickness of the strip. We check that all the mechanical quantities become null, which is in agreement with the experimental results.

\subsection{Influence of the strip geometry}
Our numerical simulations ascertain the results of the previous section with an excellent correlation: for the three permittivity models, the bending moment varies linearly with the width and thickness of the strip and is independent of its length.

The blocking force is proportional to the width and inversely proportional to the length, which is in good agreement with the results of Newbury \& Leo (2003). It is also proportional to the thickness.

\begin{figure*} [h]
\includegraphics[width=0.33\textwidth]{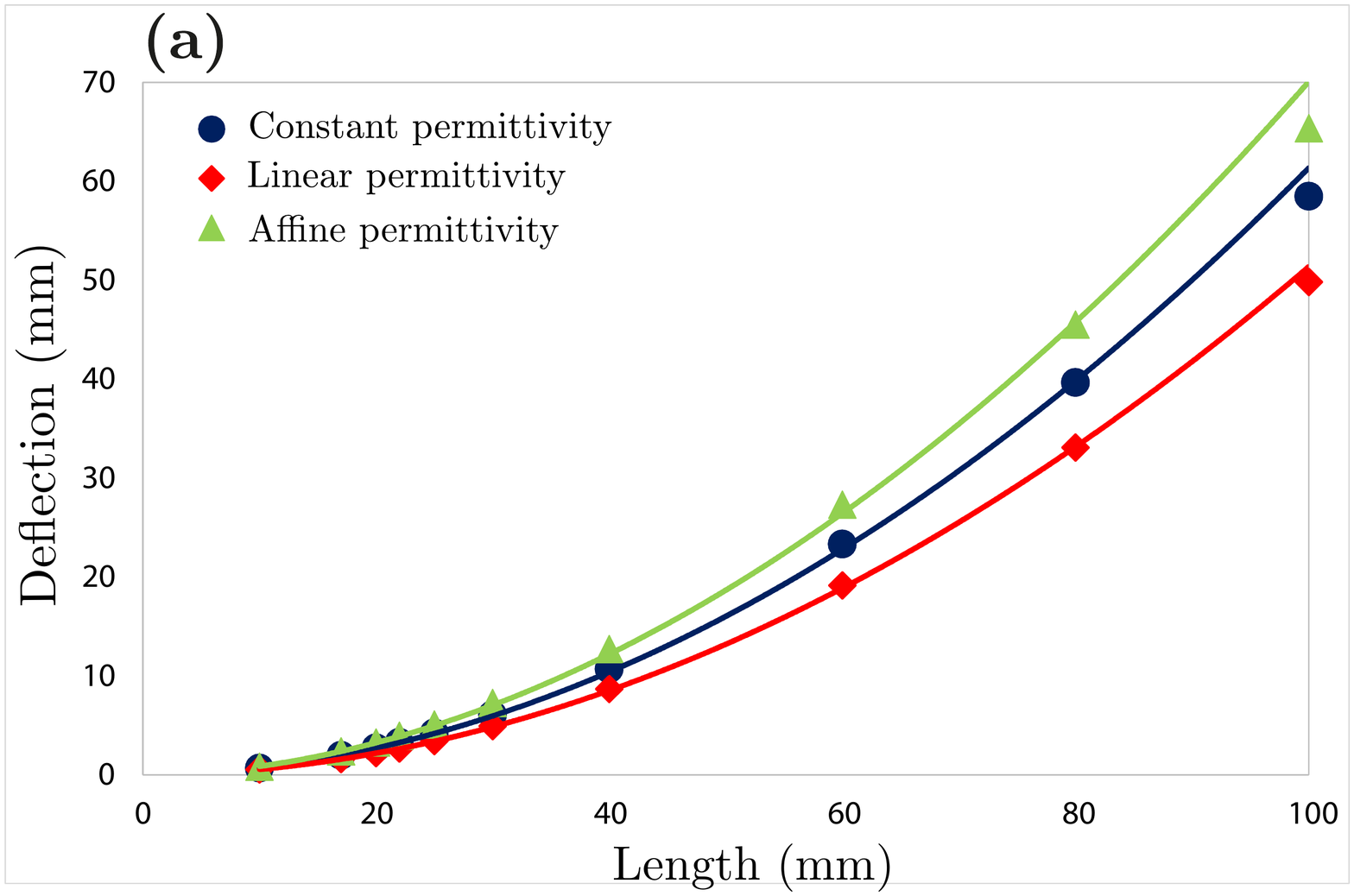}
\includegraphics[width=0.33\textwidth]{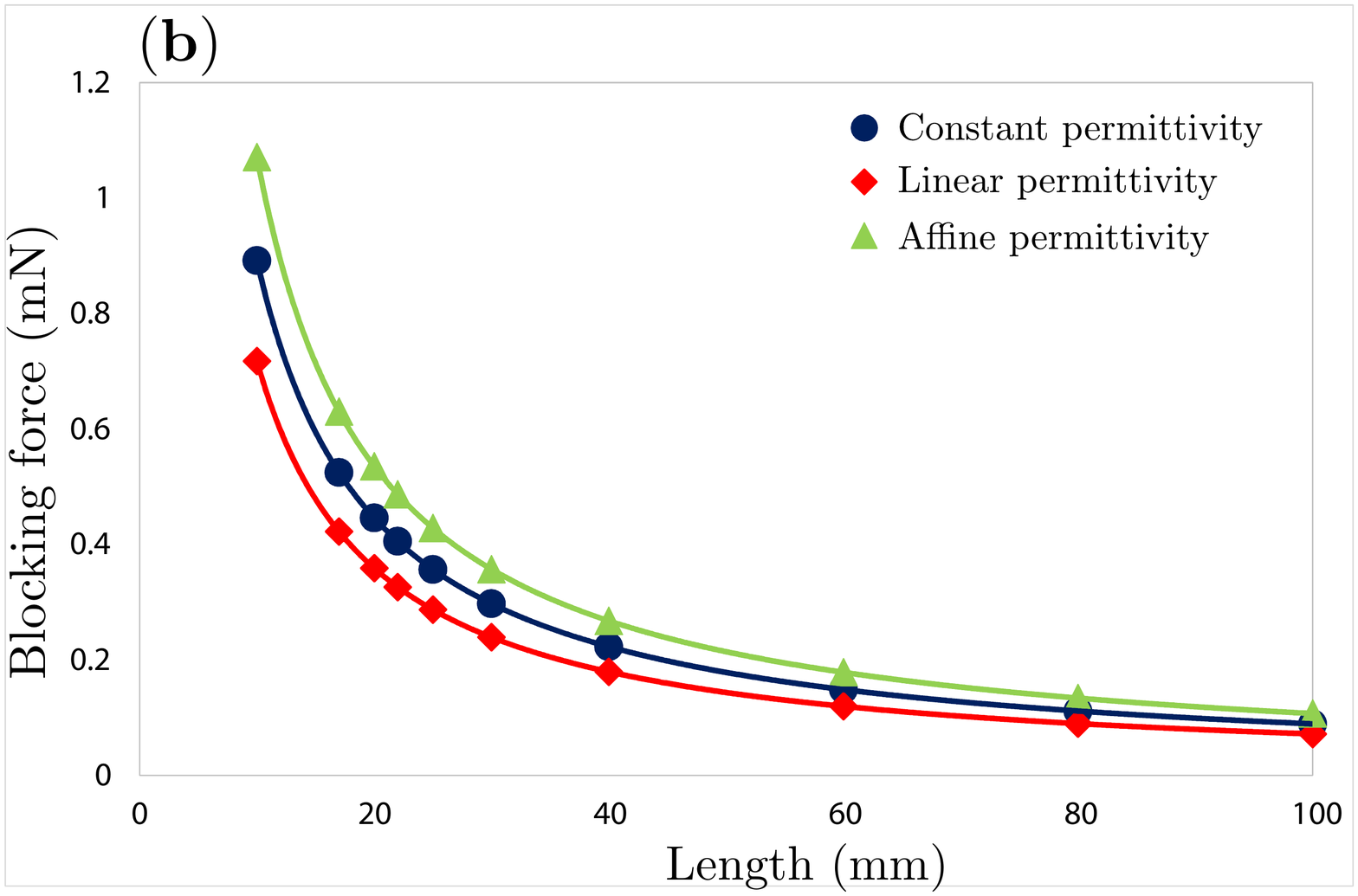}
\includegraphics[width=0.33\textwidth]{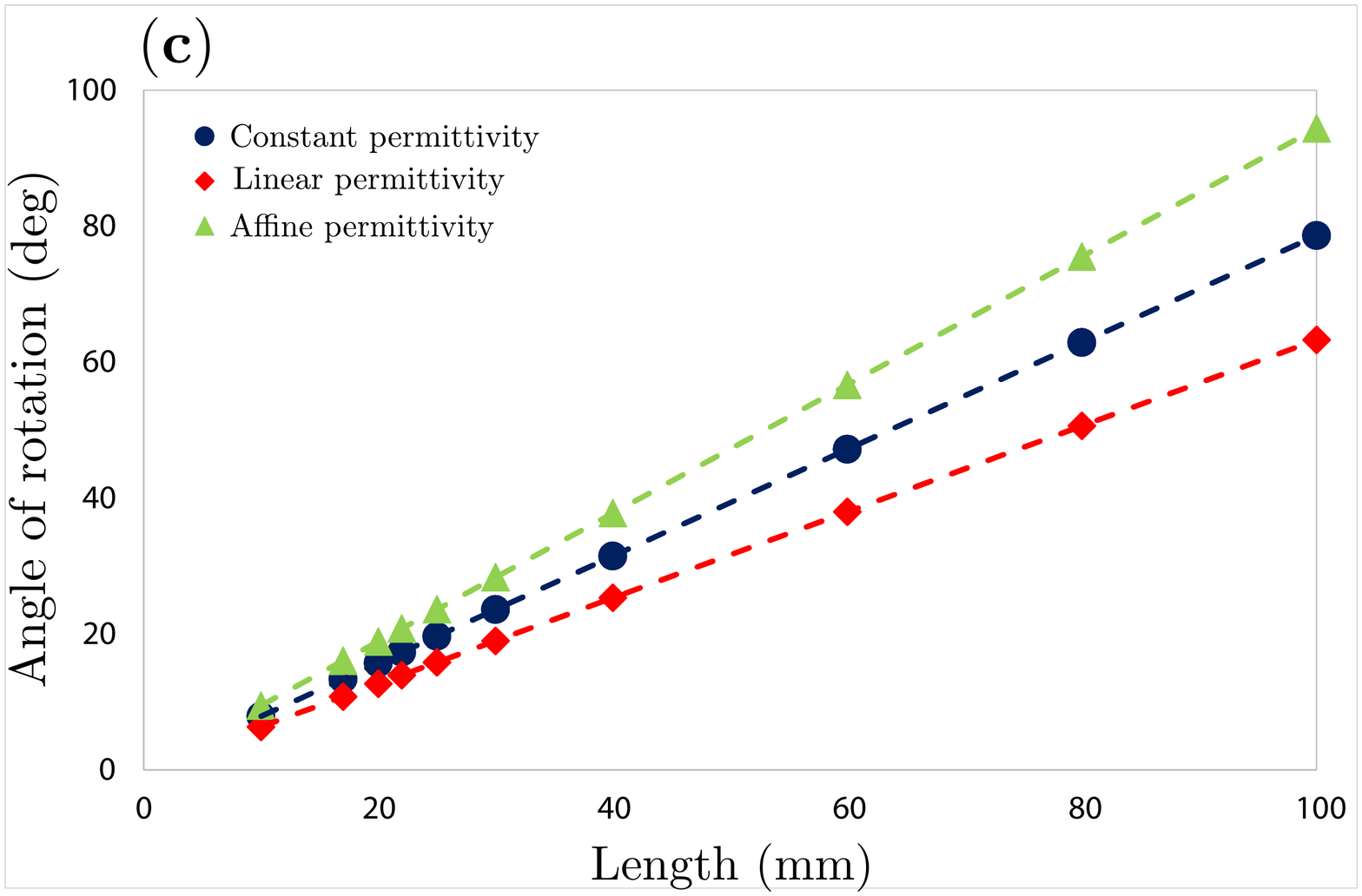}
\caption{Influence of the length: (a) on the deflection;  (b) on the blocking force;  (c) on the angle of rotation. The constant permittivity model is in blue disks, the linear one in red diamonds and the affine one in green triangles. Fitting by power laws (solid curves) or linear law (dashed curves).}
\label{fig:9}
\end{figure*}

\begin{figure*} [h]
\includegraphics[width=0.33\textwidth]{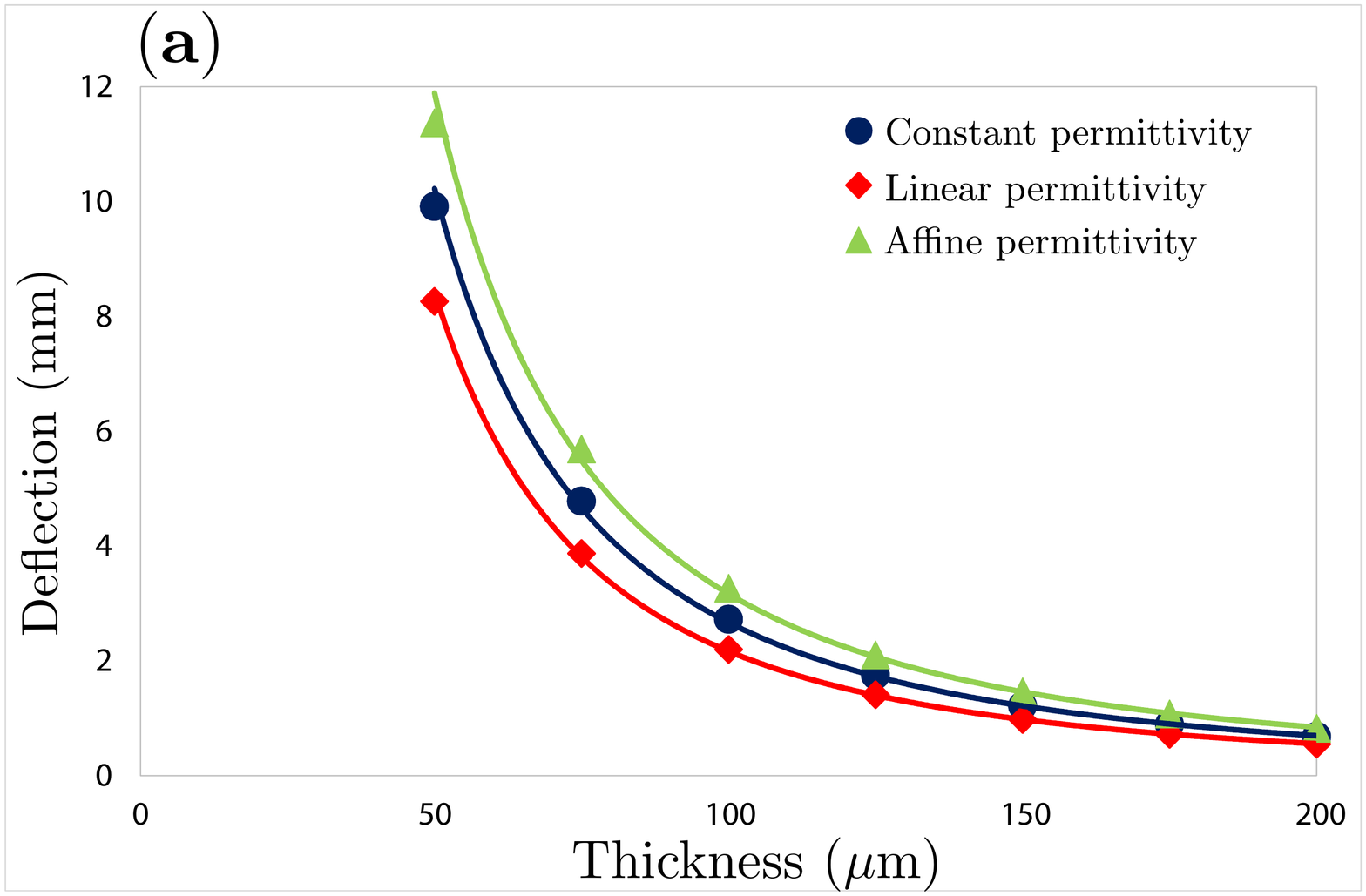}
\includegraphics[width=0.33\textwidth]{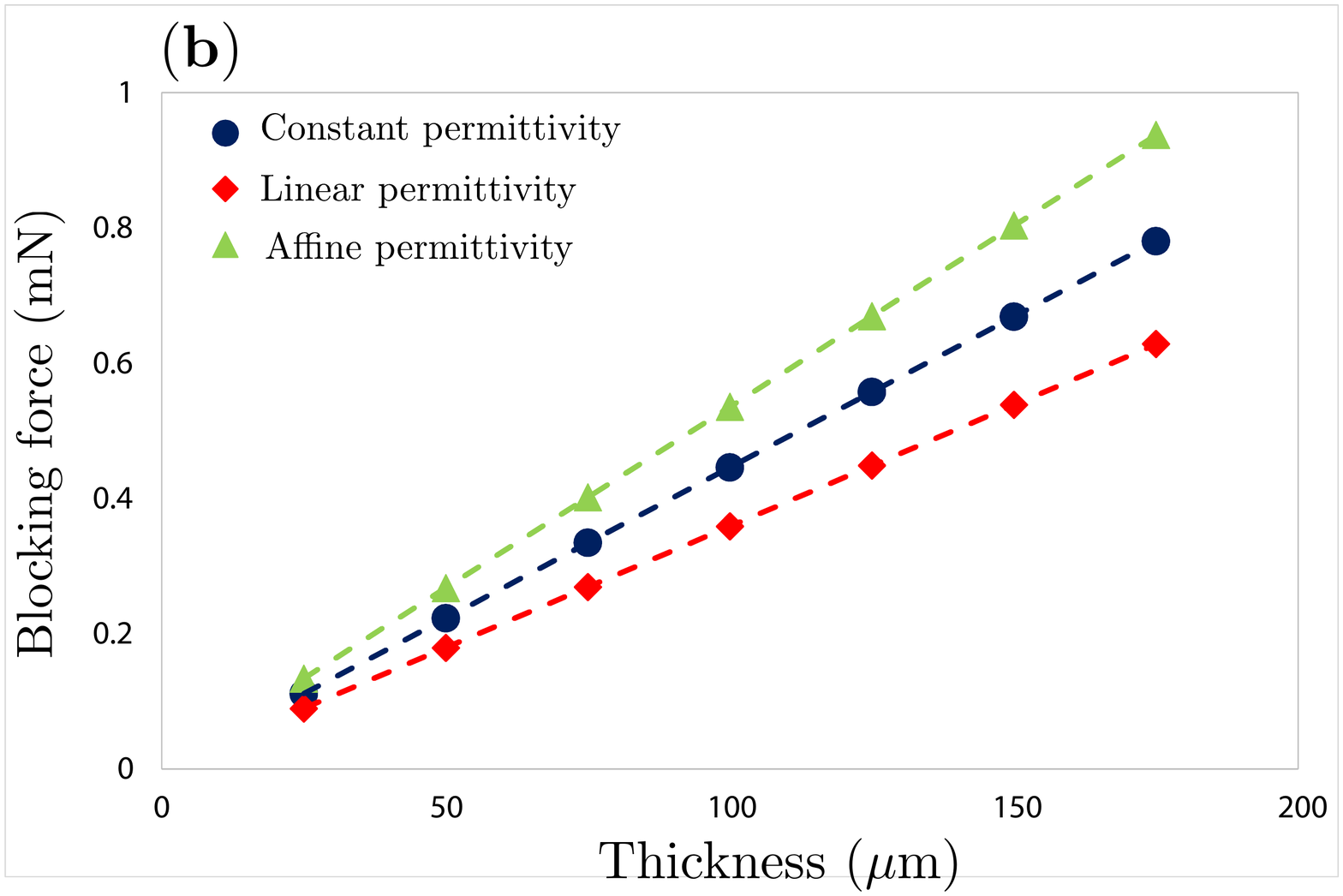}
\includegraphics[width=0.33\textwidth]{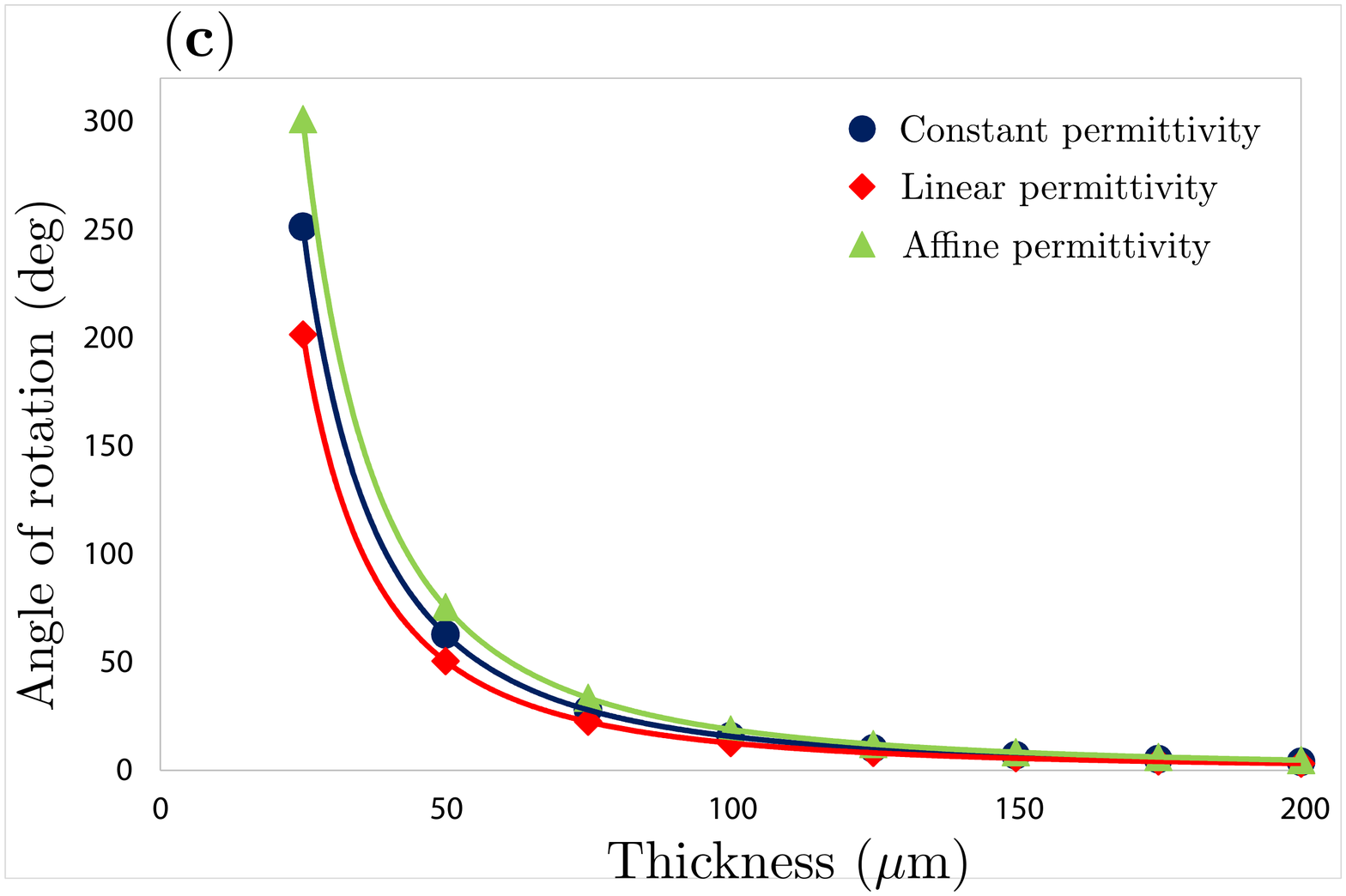}
\caption{Influence of the thickness: (a) on the deflection;  (b) on the blocking force;  (c) on the angle of rotation (same colours, marks and curves as in \fref{fig:9}).}
\label{fig:10}
\end{figure*}

The predictions of the previous section are also well fitted by the deflection in large displacements; it is independent of the width and approximately proportional to $L^{2}$: according to the chosen permittivity model, the power law that best approximates our simulations has an exponent between $1.90$ and $1.96$, and the variations of the ratio $w/L^{2}$ are less than $15\%$ in all cases (\fref{fig:9}). This is consistent with the results of Shahinpoor (1999). It varies almost like $e^{-2}$: we find an exponent between $-1.90$ and $-1.96$ according to the chosen permittivity model and the variations of the product $we^{2}$ are less than $14\%$ (\fref{fig:10}). This result is corroborated by the measurements of He et al. (2011) as well as by the simulations of Vokoun et al. (2015).

We also observe that the charge of the negative electrode $FC(e)=FB_{1}C_{moy}$ is independent of the thickness $e$ with the three permittivity models, which agrees with the results obtained by Lin et al. (2012) for a close material.

\subsection{Influence of the imposed electric potential}
\begin{figure} [h]
\includegraphics[width=0.45\textwidth]{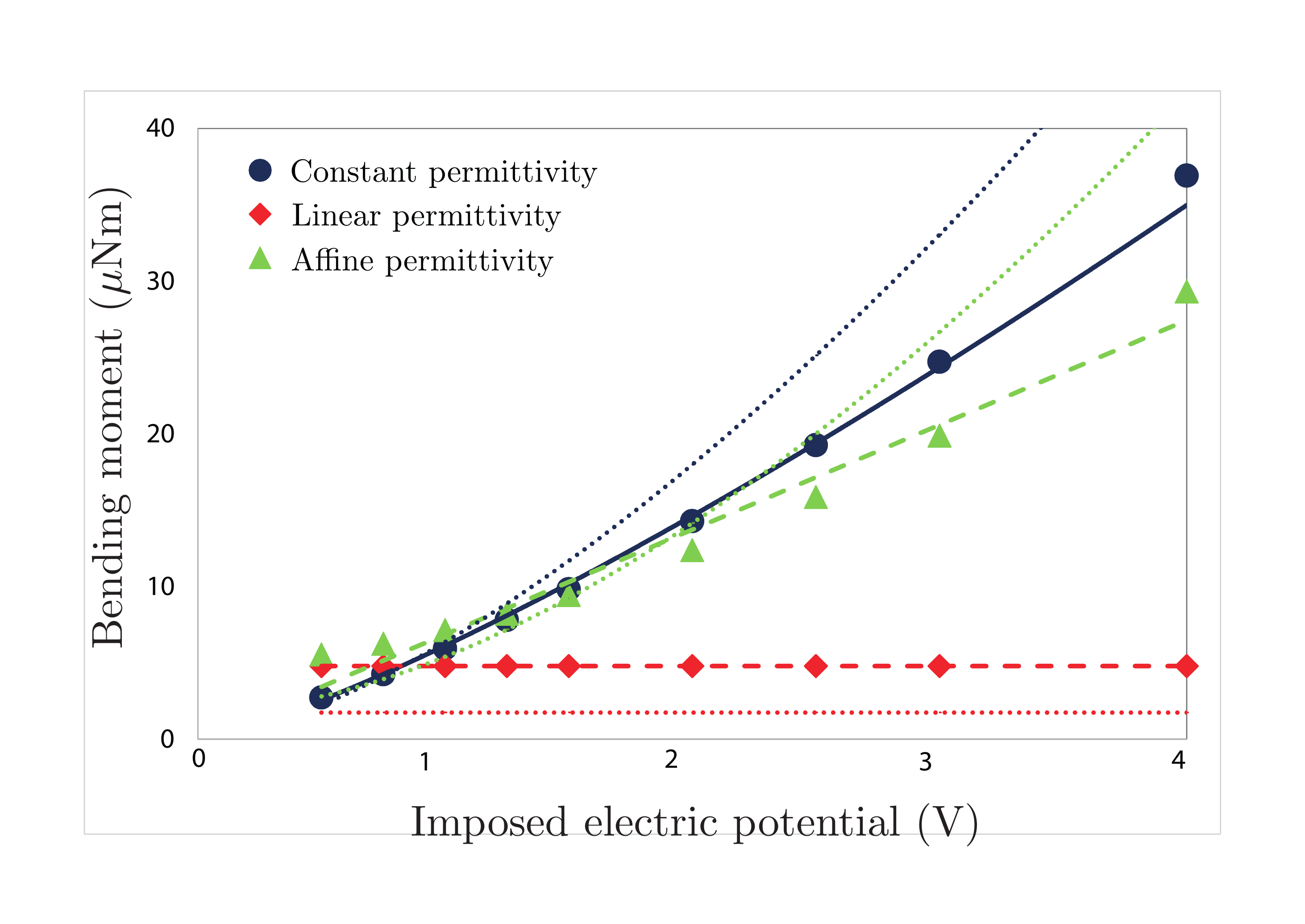}
\caption{Influence of the electric potential on the bending moment: fitting by power law (solid curve), affine laws (dashed curves) and with equations (\ref{lissage}) (dotted curves); same colours and marks as in \fref{fig:9}.}
\label{fig:11}
\end{figure}

Unlike scaling laws, the relation between the different mechanical quantities and $\varphi_{0}$ depends on the chosen permittivity model. According to equations (\ref{Meca}), the angle of rotation, the blocking force, the deflection in small displacements and the bending moment vary over the imposed potential in the same way. First, we check that the bending moment tends to $0$ when $\varphi_{0}$ tends to $0$ in all three cases using Taylor expansions. For imposed potentials close to $1~V$, we have seen that $M^{p}_{A}$ is proportional to $\varphi_{0}^{3/2}$ in the case of a constant permittivity, is almost constant if the permittivity is linear and is a complex function $f(\varphi_{0})$ in the case of an affine permittivity (\tref{t3}, \fref{fig:11}).

\begin{figure*} [h!]
\includegraphics[width=0.33\textwidth]{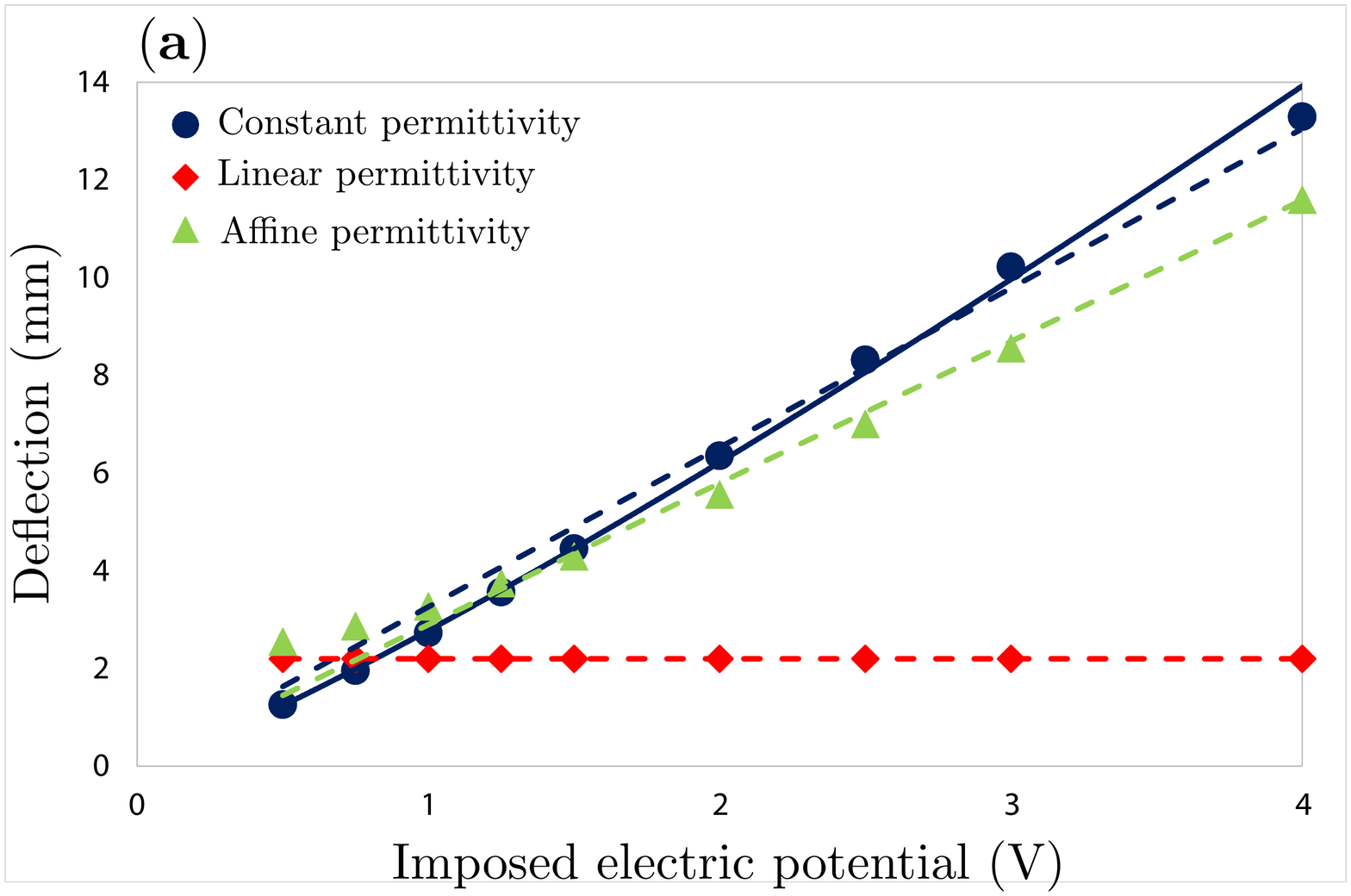}
\includegraphics[width=0.33\textwidth]{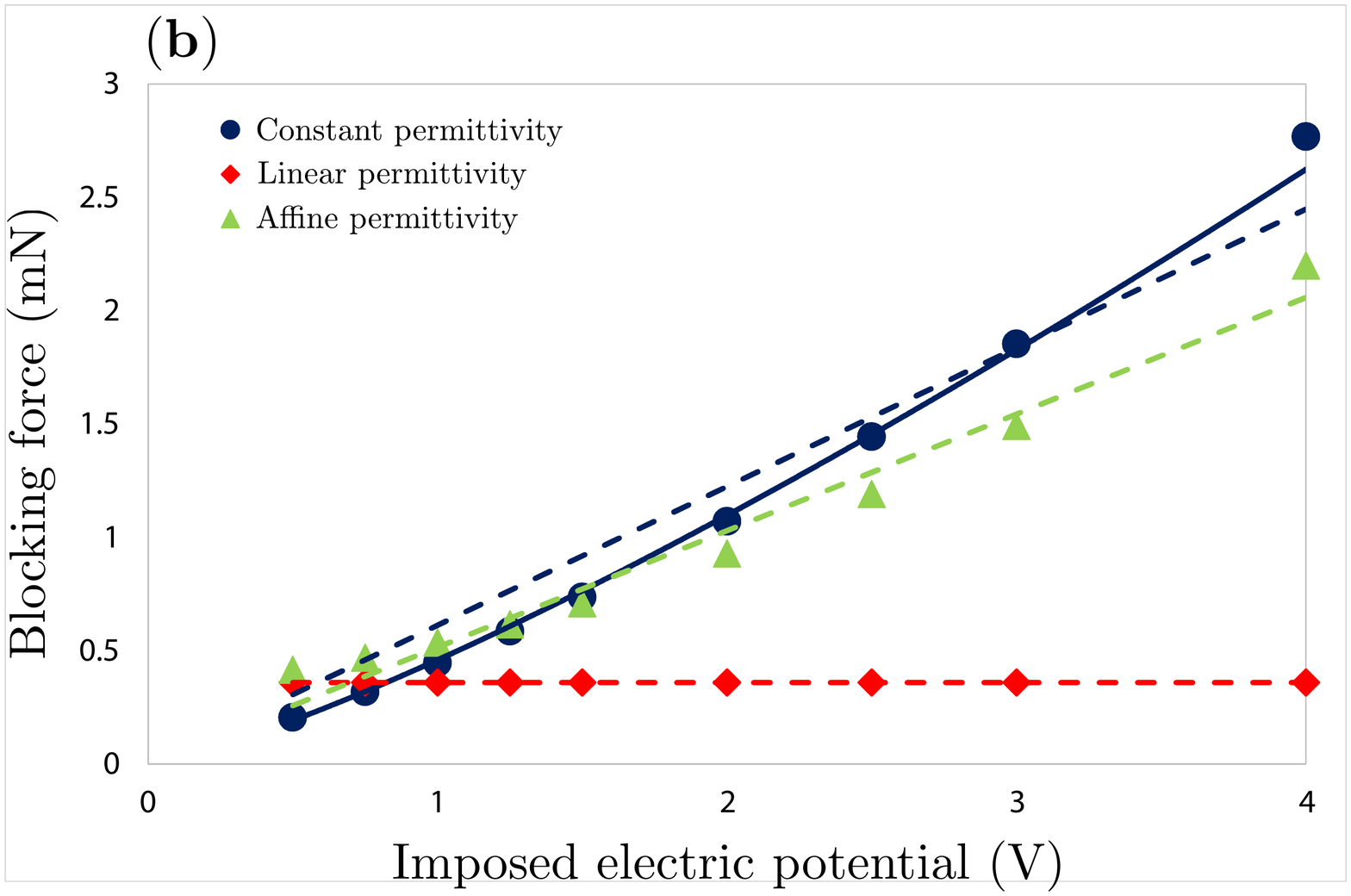}
\includegraphics[width=0.33\textwidth]{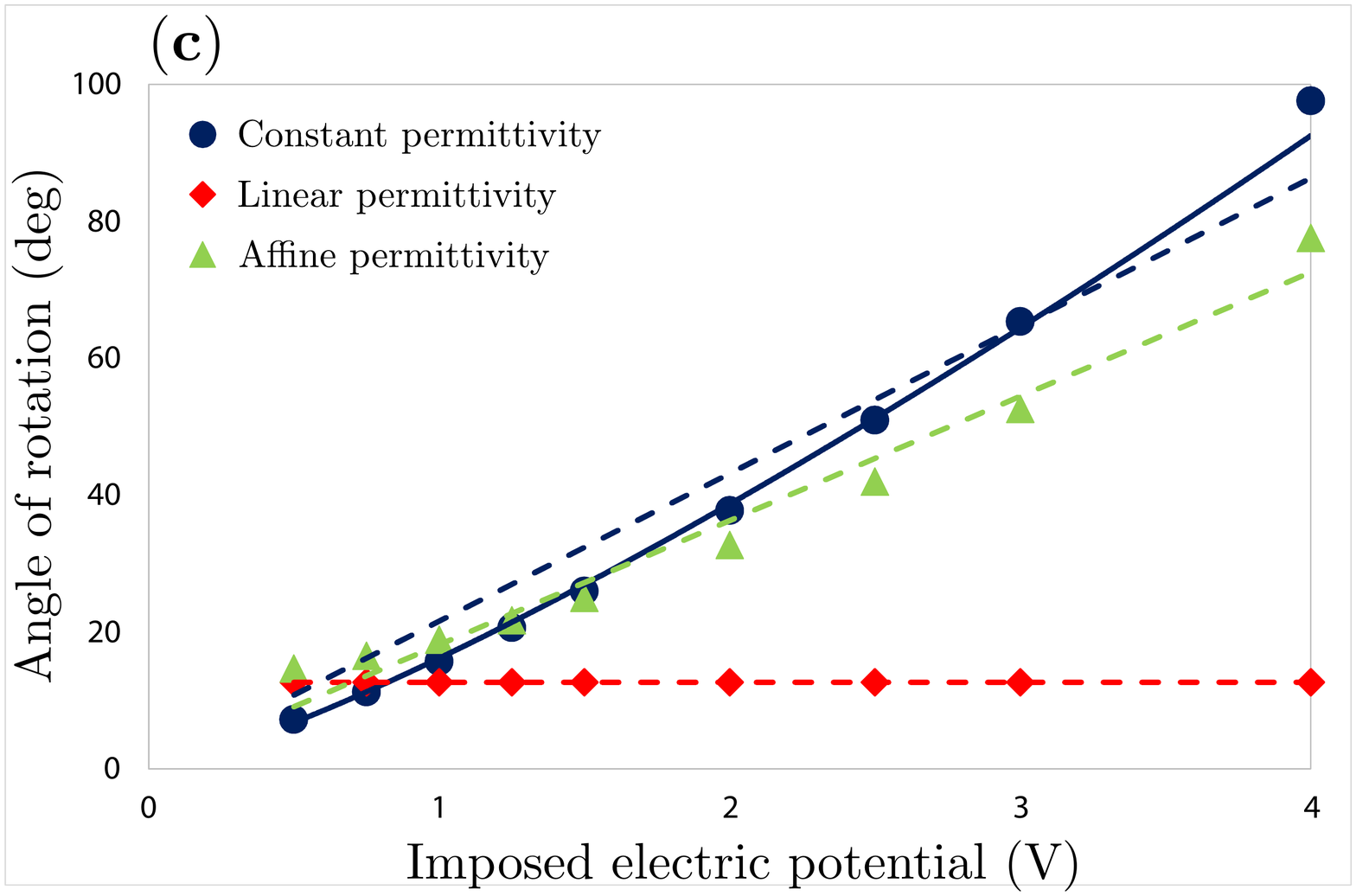}
\caption{Influence of the electric potential: (a) on the deflection;  (b) on the blocking force;  (c) on the angle of rotation (same colours marks and curves as in \fref{fig:9}).}
\label{fig:12}
\end{figure*}

Bakhtiarpour et al. (2016) observed experimentally an approximately linear relation between the deflection and the potential for actuators, and Shahinpoor et al. (1998) and Mojarrad \& Shahinpoor (1997) did the same for the sensor effect. Our simulations show a quite good correlation with a linear law if the permittivity is constant (the variation of the ratio $w/\varphi_{0}$ is about $30\%$); on the contrary, the results obtained with the other two models of permittivity do not agree with these experimental results.
More precisely, in the constant case, the curve that best fits our numerical results is a power law of exponent $1.16$ (\fref{fig:12}); this curve can hardly be distinguished from an experimental linear curve (Tixier \& Pouget 2018).
The blocking force and the angle of rotation follow approximately the same trend (good correlation with a linear law and with a power law of exponent $1.26$). This result is in good agreement with Hasani et al. (2019) experimental data, which can be well fitted by a power law of exponent $1.5$.
In the linear case, the moment is independent of the imposed potential for $\varphi_{0} \gtrsim 1$, as well as the deflection, the blocking force and the angle of rotation, which does not correspond to the experimental observations. In the affine case, the correlations with a linear law and with a power law are wrong for all the quantities.
The variation of the different quantities with the imposed electric potential is thus discriminating for the permittivity model: only a constant permittivity gives results compatible with the experimental studies.

We can evaluate the average permittivity of the strip by analogy with a capacitor of thickness $e$ and surface $Ll$ subject to a potential difference $\varphi_{0}$. We saw that the cations accumulate near the negative electrode over a thickness of $e'<<e$, and that there was a similar thickness area without cations near the positive electrode. The electric charge density is about $\rho Z^{+}=F C_{moy}(B_{1}-1)$ near the positive electrode and $\rho Z^{-}=- F C_{moy}$ near the negative one. The capacitive charge $Q$ can therefore be estimated by
\begin{equation}
Q = (\rho Z^{+}-\rho Z^{-}) \phi_{4} l L e'= F C_{moy}\phi_{4} l L e' B_{1} \,.
\end{equation}
The permittivity is then worth
\begin{equation}
\varepsilon =\frac{Q e}{L l \varphi_{0}} = \frac{F C_{moy}\phi_{4} e e'}{\varphi_{0}} B_{1} \,.
\end{equation}
In the case of constant permittivity, $B_{1}=A_{2}$ hence $\varepsilon \sim 4 \; 10^{-3} - 4 \; 10^{-2}$. In the case of a linear or affine permittivity, $B_{1} \simeq 2$ and $\varepsilon \sim \; 10^{-2}$. These values are consistent with the permittivity values derived from capacity measurements published in the literature (Nemat-Nasser 2002, Farinholt \& Leo 2004).

The hydrated IPMC can be considered as a five-layer capacitor: the electrodes of thicknesses $e_{1}$ and $e_{5}$, the central area of thickness $e_{3}$ very large compared to the previous thicknesses and of zero electric charge, and two very thin zones of the polymer of respective thicknesses $e_{2}$ and $e_{4}$ in the vicinity of the positive and negative electrodes. Each element has a permittivity $ \varepsilon_{i}$ and a capacity $C_{i} = \varepsilon_{i} \frac{L}{e_{i}}$, and the five elements are in series. The permittivity of a material being closely related to its conductivity, so here to the cation concentration, $\varepsilon_{1}= \varepsilon_{5}>> \varepsilon_{2}>> \varepsilon_{3} \gtrsim \varepsilon_{4}$. In addition $e_{3}$ is much greater than $e_{1}$, $e_{2}$, $e_{4}$ and $e_{5}$. We deduce
\begin{equation*}
\varepsilon = \frac{e}{\frac{e_{1}+e_{5}}{\varepsilon_{1}}+\frac{e_{2}}{\varepsilon_{2}}+\frac{e_{3}}{\varepsilon_{3}}+\frac{e_{4}}{\varepsilon_{4}}}
\simeq \varepsilon_{3} \,,
\end{equation*}
where $e$ denotes the total thickness. The overall permittivity of the strip subject to an electric field is therefore very close to the permittivity of the central part, which is that of the hydrated polymer without electric field. Our average permittivity values range from $5~10^{-7}~F~m^{-1}$ to $10^{-4}~F~m^{-1}$ depending on the permittivity model and are thus compatible with the permittivity values measured by Deng \& Mauritz (1992) and Wang et al. (2014), which range from $10^{-7}~F~m^{-1}$ to $5~10^{-3}~F~m^{-1}$.

\section{Conclusion}
We studied the flexion of an ionic polymer strip using a model based on the thermomechanical of continuous media that we had previously developed. The strip, clamped at one of its ends, is subject to a continuous potential difference applied between its two faces (static case). The other end is either free or blocked by a shear force. The mechanical quantities (deflection, blocking force and angle of rotation) were determined using a beam model in large displacements. The material chosen to perform the simulations is Nafion$^{\mbox{\scriptsize{\textregistered}}}Li^{+}$. Three models of permittivity (constant, linear and affine functions of cation concentration) are examined. 
The permittivity values we used are in good agreement with dielectric spectroscopy and electrical impedance measurements, but significantly lower than those deduced from capacity measurements.
The resolution of the equations of our model enabled us to plot the cation concentration, pressure, electric potential and displacement profiles over the thickness of the strip. These quantities are almost constant in the central part, but vary drastically in the vicinity of the electrodes, which is characteristic of a conductive material. The scaling laws obtained for the deflection and the blocking force are in good agreement with the experimental data published in the literature: in particular, the deflection varies as the square of the strip length and is inversely proportional to the square of its thickness; the blocking force is proportional to the width and the thickness and it is inversely proportional to the length.
The variation of the mechanical quantities with the imposed electric potential depends on the chosen permittivity model; only the constant permittivity model provides results compatible with the experimental data and will therefore be retained for further works.
We now plan to apply our model to other materials close to the Nafion$^{\mbox{\scriptsize{\textregistered}}}$ and to study other configurations such as a strip clamped at its two ends.

\section*{Notations}
$k=1,2,3,4,i$ subscripts respectively represent cations, solvent, solid, solution (water and cations) and interface; quantities without subscript refer to the whole material. Superscript $^{0}$ denotes a local quantity; the lack of superscript indicates average quantity at the macroscopic scale. Superscript $^{T}$ indicates the transpose of a second-rank tensor. Overlined letters denote dimensionless quantities.

\begin{description}
\item $A_{i}$, $B_{i}$: dimensionless constants;

\item $C$, $C_{moy}$: cations molar concentrations (relative to the liquid phase);

\item $\mathsf{D}$: mass diffusion coefficient of the cations in the liquid phase;

\item $\vec{D}$: electric displacement field;

\item $e$: half-thickness of the strip;

\item $E$, $G$, $\lambda$: Young's and shear modulus, first Lam\'{e} constant;

\item $\vec{E}$: electric field;

\item $F=96487\;C\;mol^{-1}$: Faraday's constant ;

\item $\vec{F^{p}}$ ($\vec{F^{p}_{s}}$): blocking force on large (small) displacements;

\item $\vec{I}$: current density vector;

\item $I^{p}$: moment of inertia;

\item $K$: intrinsic permeability of the solid phase;

\item $l$: half-width of the strip;

\item $L$: length of the strip;

\item $M_{k}$: molar mass of component $k$;

\item $M_{eq}$: equivalent weight (weight of polymer per mole of sulfonate groups);

\item $\vec{M^{p}}$ ($\vec{M_{A}^{p}}$): bending moment;

\item $\vec{n}$ ($\overline{n}$): normal vector (coordinate) to the beam;

\item $p$: pressure;

\item $\overrightarrow{p^{p}}$: distributed electric force

\item $\vec{Q}$: heat flux;

\item $R=8,314\;J\;K^{-1}$: gaz constant;

\item $R^{p}$: radius of curvature of the beam;

\item $s$ ($\overline{s}$): curvilinear abscissa along the beam at rest (deformed)

\item $S$: entropy density;

\item $T$: absolute temperature;

\item $\vec{u}$: displacement vector;

\item $U$: internal energy density;

\item $v_{k}$: partial molar volume of component $k$ (relative to the
liquid phase);

\item $\vec{V}$ ($\vec{V_{k}}$): velocity;

\item $w$ ($w_{s}$): deflection of the beam on large (small) displacements;

\item $Z$ ($Z_{k}$): total electric charge per unit of mass;

\item $\varepsilon$ ($\varepsilon^0$, $\varepsilon_{moy}$): permittivity (average permittivity);

\item $\utilde{\epsilon}$: strain tensor;

\item $\eta_{2}$: dynamic viscosity of water;

\item $\theta$: angle of rotation of a beam cross section;

\item $\lambda_{v}$, $\mu_{v}$: viscoelastic coefficients;

\item $\mu_{k}$: mass chemical potential;

\item $\rho $ ($\rho _{k}$): mass density relative to the volume of the whole material;

\item $\rho _{k}^{0}$: mass density relative to the volume of the phase;

\item $\utilde{\sigma}$ ($\utilde{\sigma_{k}}$), $\utilde{\sigma^{e}}$ : stress tensor, equilibrium stress tensor;

\item $\phi_{k}$: volume fraction of phase $k$;

\item $\varphi$ ($\varphi_{0}$): electric potential (imposed electric potential);

\end{description}

\section*{References}
\begin{harvard}
\item[] Aureli, M. and Prince, C. and Porfiri, M. and Peterson, S.D. 2010 \textit{Smart Materials and Structures} \textbf{19}, 015003.

\item[] Bahramzadeh, Y. and Shahinpoor, M. 2014 \textit{Soft Robotics} \textbf{1}(1), 38--52.

\item[] Bakhtiarpour, P. and Parvizi, A. and Müller, M. and Shahinpoor, M. and Marti, O. and Amirkhani, M. 2016 \textit{Smart Materials and Structures} \textbf{25}, 015008 doi:10.1088/0964-1726/25/1/015008.

\item[] Bar-Cohen, Y. 2005 \textit{WIT Transactions on State of the Art in Science and Engineering} \textbf{20}, 66--81.

\item[] Barclay Satterfield, M. and Benziger, J. B. 2009 \textit{J. Polym. Sci. Pol. Phys.} \textbf{47}(1), 11--24.

\item[] Bauer, F. and Denneler, S. and Willert-Porada, M. 2005 \textit{Journal of Polymer Science part B : Polymer Physics} \textbf{43}(7), 786--795.

\item[] Biot, M.A. 1977 \textit{International Journal of Solids and Structures} \textbf{13}, 579--597.

\item[] Bluhm, J. and Serdas, S. and  Schröder, J. 2016 \textit{Archive of Applied Mechanics} \textbf{86}, 3--19.

\item[] Brunetto, P. and Fortuna, L. and Graziani, L. and
Strazzeri, S. 2008 \textit{Smart Materials and Structures} \textbf{17}, 025029.

\item[] Cappadonia, M. and Erning, J. and Stimming, U. 1994 \textit{Journal of Electroanalytical Chemistry} \textbf{376}(1), 189--193.

\item[] Cha, Y. and Shen, L. and Porﬁri M. 2013 \textit{Smart Materials and Structures} \textbf{22}, 055027.

\item[] Chabé, J. 2008 Etude des interactions moléculaires polymère-eau lors de l'hydratation de la membrane Nafion, électrolyte de référence de la pile à combustible. PhD thesis Université Joseph Fourier Grenoble I \textit{http://tel.archives-ouvertes.fr/docs/00/28/59/99/PDF/THESE\_JCS.pdf}.

\item[] Chen, Z. 2017 \textit{Robotics and Biomimetics} \textbf{4}(24), https://doi.org/10.1186/s40638-017-0081-3

\item[] Chikhaoui, M.T. and  Benouhiba, A. and Rougeot, P. and Rabenorosoa, K. and Ouisse, M. and Andreff, N. 2018 \textit{Annals of Biomedical Engineering} \textbf{16}(10), 1511--1521. https://doi.org/10.1007/s10439-018-2038-2.

\item[] Collette, F. 2008 Vieillissement hygrothermique du Nafion. PhD Thesis Ecole Nationale Supérieure des Arts et Métiers \textit{$http://tel.archives-ouvertes.fr/docs/00/35/48/47/PDF/These\_Floraine$ $\_COLLETTE\_27112008.pdf$}.

\item[] Coussy, O. 1995 \textit{Mechanics of porous continua.} Wiley Chichester.

\item[] de Groot, S. R. and Mazur, P. 1962 \textit{Non-equilibrium thermodynamics.} North-Holland publishing company Amsterdam

\item[] Deng, Z.D. and Mauritz, K.A. 1992 \textit{Macromolecules} \textbf{25}(10), 2739--2745.

\item[] Deole, U. and Lumia, R. and Shahinpoor, M. and Bermudez, M. 2008 \textit{Journal of Micro-Nano Mechatronics} \textbf{4}, 95--102.

\item[] Fang, B.K. and Ju, M.S. and Lin, C.C.K. 2007 \textit{Sensors and Actuators A} \textbf{137}(2), 321--329.

\item[] Farinholt, K. and Leo, D.J. 2004 \textit{Mechanics of Materials} \textbf{36}(5), 421--433.

\item[] Farinholt, K. and Pedrazas, N.A. and Schluneker, D.M. and Burt, D.W. and Farrar, C.R. 2009 \textit{Journal of Intelligent Material Systems and Structures} \textbf{20}(5), 633--642.

\item[] Festin, N. and Plesse, C. and Pirim, P. and Chevrot, C. and Vidal, F. 2014 \textit{Sensors and Actuators B: Chemical} \textbf{193}, 82--88.

\item[] Gierke, T.D. and Munn, G.E. and Wilson, F.C. 1981 \textit{Journal of Polymer Science : Polymer Physics Edition} \textbf{19}(11), 1687--1704.

\item[] Hasani, M. and Alaei, A. and Mousavi, M. S. S. and Esmaeili, E. and Kolahdouz, M. and Naeini, V. F. and Masnadi-Shirazi, M. 2019 \textit{Journal of Micromechanics and Microengineering} \textbf{29}, 085008 https://doi.org/10.1088/1361-6439/ab272c.

\item[] He, Qingsong and Yu, Min and Song, Linlin and Ding, Haitao and Zhang, Xiaoqing and Dai, Zhendong 2011 \textit{Journal of Bionic Engineering} \textbf{8}, 77--85.

\item[] Ishii, M. and Hibiki, T. 2006 \textit{Thermo-fluid dynamics of two-phase flow.} Springer New-York.

\item[] Jean-Mistral, C. and Basrour, S. and Chaillout, J.J. 2010 \textit{Smart Materials and Structures} \textbf{19}, 085012.

\item[] Jo, C. and Pugal, D. and Oh, I. and Kim, K.J. and
Asaka, K. 2013 \textit{Progress in Polymer Science} \textbf{38}, 1037--1066

\item[] Lakshminarayanaiah, N. 1969 \textit{Transport Phenomena in Membranes.} Academic Press New-York.

\item[] Lin, J. and Liu, Y. and Zhang, Q.M. 2012 \textit{Macromolecules} \textbf{45}(4), 2050--2056.

\item[] Matysek, M. and Lotz, P. and Schlaak H.F. 2009 \textit{Proc. SPIE 7287, Electroactive Polymer Actuators and Devices (EAPAD)} \textbf{7287}, 72871D doi: 10.1117/12.819217.

\item[] Mojarrad, M. and Shahinpoor, M. 1997 \textit{Proc. SPIE} \textbf{3042}, 52--60.

\item[] Nardinocchi, P. and Pezzulla, M. and Placidi, L. 2011 \textit{Journal of Intelligent Material Systems and Structures}\textbf{22}(16), 1887--1897.

\item[] Nemat-Nasser, S. 2002 \textit{Journal of Applied Physics} \textbf{92}(5,) 2899--2915.

\item[] Nemat-Nasser, S. and Li, J. 2000 \textit{Journal of Applied Physics} \textbf{87}(7), 3321--3331.

\item[] Newbury, K.M. 2002 Characterization, modeling and control of ionic-polymer transducers. PhD thesis Faculty of the Virginia Polytechnic Institute and State University Blacksburg, Virginia.

\item[] Newbury, K.M. and Leo, D.J. 2002 \textit{Journal of Intelligent Material Systems and Structures} \textbf{13}(1), 51--60.

\item[] Newbury, K.M. and Leo, D.J. 2003 \textit{Journal of Intelligent Material Systems and Structures} \textbf{14}(6), 343--357.

\item[] Nguyen, N.T. and Dobashi, Y. and Soyer, C. and Plesse, C. and Nguyen, G.T.M. and Vidal, F. and Cattan, E. and Grondel, S. and Madden, J.D.W. 2018 \textit{Smart Materials and Structures} \textbf{27}, 115032.

\item[] Park, I. and Kim, S.M. and Pugal, D. and Huang, L. and Tam-Chang, S.W. and Kim K.J. 2010 \textit{Applied Physics Letters} \textbf{96}(4), 043301.

\item[] Pugal, D. and Jung, K. and Aabloob, A. and Kim, K.J. 2010 \textit{Polymer International} \textbf{59}, 279--289.

\item[] Shahinpoor, M. 1999 \textit{Proc. of SPIE} \textbf{3669}, 109--121.

\item[] Shahinpoor, M. and Bar-Cohen, Y. and Simpson, J.O. and Smith, J. 1998 \textit{Smart Materials and Structures} \textbf{7}(6), R15--R30.

\item[] Shahinpoor, M. and Kim, K.J. 2001 \textit{Smart Materials and Structures} \textbf{10}(4), 819--833.

\item[] Silberstein, M. N. and Boyce, M. C. 2010 \textit{J. Power Sources} \textbf{195}(17), 5692--5706.

\item[] Tiwari, R. and Kim, K. J. and  Kim, S. M. 2008 \textit{Smart Structures and Systems} \textbf{4}(5), 549. DOI: 10.12989/sss.2008.4.5.549.

\item[] Tixier, M. and Pouget, J. 2014 \textit{Continuum Mechanics and Thermodynamics} \textbf{26}(4), 465--481.

\item[] Tixier, M. and Pouget, J. 2016 \textit{Continuum Mechanics and Thermodynamics} \textbf{28}(4), 1071--1091.

\item[] Tixier, M. and Pouget, J. 2018 \textit{Chapter 39: Modelling of an Ionic Electroactive Polymer by the Thermodynamics of Linear Irreversible Processes} in Generalized Models and Non Classical Mechanical Approaches in Complex Materials Springer Berlin.

\item[] Vokoun, D. and Qingsong, He and Heller, L. and Min, Y. and Zhen, D.D. 2015 \textit{Journal of Bionic Engineering} \textbf{12}(1), 142--151.

\item[] Wallmersperger, T. and Horstmann, A. and Kröplin, B. and Leo, D.J. 2009 \textit{Journal of Intelligent Material Systems and Structures} \textbf{20}, 741--750.

\item[] Wang, Y. and Zhu, Z. and Chen, H. and Luo, B. and Chang, L. and Wang, Y. and Li, D. 2014 \textit{Smart Materials and Structures} \textbf{23}(125015). doi:10.1088/0964-1726/23/12/125015.

\end{harvard}

\end{document}